\documentclass[usenatbib,usegraphicx]{mn2e}
\usepackage{amssymb}

\input epsf

\def\Lya{Ly$\alpha$}

\def\HI{\hbox{H~$\scriptstyle\rm I\ $}}

\def\HII{\hbox{H~$\scriptstyle\rm II\ $}}

\def\kms{\,{\rm {km\, s^{-1}}}}
\def\msun{{M_\odot}}

\def\kmsmpc{\,{\rm km\,s^{-1}\,Mpc^{-1}}}

\def\ltsima{$\; \buildrel < \over \sim \;$}
\def\lsim{\lower.5ex\hbox{\ltsima}}
\def\gtsima{$\; \buildrel > \over \sim \;$}
\def\gsim{\lower.5ex\hbox{\gtsima}}

\def\spose#1{\hbox to 0pt{#1\hss}}
\def\lta{\mathrel{\spose{\lower 3pt\hbox{$\mathchar"218$}}
     \raise 2.0pt\hbox{$\mathchar"13C$}}}
\def\gta{\mathrel{\spose{\lower 3pt\hbox{$\mathchar"218$}}
     \raise 2.0pt\hbox{$\mathchar"13E$}}}

\journal{Preprint-00}

\title{The impact of Lyman series photons on the intergalactic medium
  during the Epoch of Reionisation}

\author[A. Meiksin]{Avery Meiksin \\
SUPA\thanks{Scottish Universities Physics Alliance},
Institute for Astronomy, University of Edinburgh,
Blackford Hill, Edinburgh\ EH9\ 3HJ, UK}

\begin{document}

\maketitle

\begin{abstract}
  The role of higher order Lyman series photons on the 21cm absorption
  or emission signature of the IGM against the background CMB during
  the Epoch of Reionisation is examined. It is shown that, taking into
  account the diminishing resonance line scattering cross section with
  increasing Lyman order, a non-negligible net scattering rate of
  higher order Lyman photons is expected. The resulting radiative
  cascades will substantially enhance the number density of \Lya\
  photons near a radiation source. It is also shown that the higher
  order Lyman series photons are able to collisionally heat the IGM by
  amounts of tens to hundreds of degrees kelvin. The possibility that
  the Wouthuysen-Field effect may be suppressed by the presence of
  dust near a galaxy is discussed, and it is shown that the higher
  order Lyman series photons would still induce the effect, but with a
  somewhat reduced 21cm radiation efficiency. It is also demonstrated
  that extended low surface brightness emission line halos will be
  produced from radiative cascades following the scattering of higher
  order Lyman series photons. These halos would provide a unique means
  of confirming that reionisation source candidates were surrounded by
  an IGM that was still largely neutral on large scales.
\end{abstract}


\section{Introduction}

The detection of the first sources of light in the Universe through
their induced intergalactic \HI\ 21cm signature before the Epoch of
Reionisation (EoR) \citep{1979MNRAS.188..791H, 1990MNRAS.247..510S,
  MMR97} may be realised in the near to not very distant future with
the advent of a new generation of metre-wavelength scale radio
telescopes, such as the LOw Frequency Array
(LOFAR)\footnote{www.lofar.org}, the Murchison Widefield Array (MWA)
\footnote{www.haystack.mit.edu/ast/arrays/mwa}, the Primeval Structure
Telescope/21 Centimeter Array (PaST/21CMA)
\footnote{web.phys.cmu.edu/$\sim$past}, the Precision Array to Probe
EoR (PAPER)\footnote{astro.berkeley.edu/$\sim$dbacker/eor}, and a
possible Square Kilometre Array (SKA)\footnote{www.skatelescope.org}.
Reviews of this rapidly growing area are provided by
\citet{2006ARA&A..44..415F} and \citet{2006PhR...433..181F}. Central
to the detection is the decoupling of the spin temperature of the
neutral hydrogen from that of the Cosmic Microwave Background (CMB).
Three possible decoupling mechanisms exist:\ coupling the energy
levels to a nearby bright radio source \citep{1969ApJ...157.1055B},
establishing thermal equilibrium with the gas through collisions by
other hydrogen atoms and electrons \citep{1958PROCIRE.46..240F}, and
coupling the energy levels to Lyman resonance line radiation through
the Wouthuysen-Field (W-F) effect
\citep{1952AJ.....57R..31W,1958PROCIRE.46..240F}, which will normally
quickly establish thermal equilibrium between the spin state and the
scattering gas. At typical intergalactic densities at redshifts
$z<17$, where the signal may be detected by the newly developed or
planned radio facilities, coupling through the scattering of Lyman
resonance line photons is expected to be the dominant mechanism.

Whilst the discussion of the W-F mechanism has normally been confined
to the context of hydrogen \Lya\ photons, the contribution of higher
order Lyman resonance line photons has received recent attention
\citep{2005ApJ...626....1B, 2006MNRAS.367..259H,
  2006MNRAS.367.1057P}. Noting that the higher order photons will
scatter only a few times before the excited atom decays through an
alternative channel, it was concluded that direct collisions by higher
order Lyman series photons would always be negligible compared with
the \Lya\ collision rate of photons emitted directly by the source. At
the same time, the \Lya\ photons produced in radiative cascades
following the scattering of higher order Lyman series photons would
provide a substantial boost to the overall scattering rate. The amount
of the boost would depend on the maximum upper principal quantum
number $n$ ($n>3$) \footnote{Ly$\beta$ photons cannot decay to
  Ly$\alpha$, but will decay to H$\alpha$ followed by two-photon
  emission.}, of the photons that are able to reach a given distance
from the source without redshifting into the the resonance frequency
of the next lower Lyman order.

In this paper, it is shown that the increasing mean free path through
the IGM of higher order Lyman resonance line photons will result in a
total scattering rate of Ly-$n$ photons a few percent of that of \Lya\
for sufficiently large upper principal quantum number $n$, much
exceeding previous estimates. The enhancement of the \Lya\ scattering
rate by \Lya\ photons produced in radiative cascades is found to boost
the \Lya\ scattering rate by up to 30 per cent. more than previous
estimates. It is also shown that since the higher order photons will
not scatter sufficiently to establish thermal equilibrium with the
gas, these photons may provide a non-negligible contribution to the
heating of the IGM through collisional heating. The heating may be
sufficient to raise the IGM temperature in the vicinity of a source
above the CMB temperature, resulting in an emission signature against
the CMB rather than absorption. Another consequence of the scattering
of higher order Lyman series photons is the production of secondary
emission lines, such as the Balmer and Paschen series, in radiative
cascades in the neutral hydrogen surrounding the source. The emission
lines would appear in the infra-red, and the emitting regions would
subtend large angles on the sky, arising in spatially extended
regions. Although the lines are individually weak, cross-correlations
of the measurements of the expected lines may provide a detectable
signal. The discovery of the emission line halos would provide a means
of confirming that high redshift reionisation candidate sources were
in fact embedded in an IGM that was still largely neutral on large
scales.

Unless stated otherwise, a flat cosmology is assumed in this paper
with a total mass density ratio to the Einstein-deSitter density of
$\Omega_m=0.3$, a vacuum energy contribution $\Omega_v=0.7$, a baryon
density $\Omega_bh^2=0.02$, and a Hubble constant of $H_0=100h\kmsmpc$
with $h=0.7$.

\section{The scattering of intergalactic Lyman resonance line
  photons}
\label{sec:IGM}

The optical depth through a homogeneous and isotropic expanding IGM of
a Ly-$n$ photon (with upper state principal quantum number $n$)
emitted by a source at redshift $z_S$ and received at redshift $z$ at
frequency $\nu>\nu_{lu}$, where $\nu_{lu}$ is the resonance line
frequency, is
\begin{equation}
\tau_{\nu} = \sigma_n\int_z^{z_S}\,dz^\prime \frac{dl_p}{dz^\prime}
n_l(z^\prime)\varphi_V\left(a_n,\nu\frac{1+z^\prime}{1+z}\right),
\label{eq:taunu}
\end{equation}
where $n_l(z^\prime)$ is the number density of scattering atoms in the
lower level at epoch $z^\prime$, $\sigma_n=(\pi e^2/m_ec)
f_{lu}\simeq0.02643f_{lu}\,{\rm cm^2\,Hz}$ is the total resonance line
cross section, $f_{lu}$ is the upward oscillator strength,
$\varphi_V(a_n,\nu)$ is the Voigt line profile normalized to $\int\,
d\nu\varphi_V(a_n,\nu)=1$, $a_n$ is the ratio of the decay rate to the
Doppler width $\Delta\nu_D=\nu_{lu} b/c$, where $b=(2k_{\rm B}T/m_{\rm
  H})^{1/2}$ is the Doppler parameter of the gas at temperature $T$
and $c$ is the speed of light, and $l_p$ is the proper path
length.\footnote{For low Lyman orders, the optical depth is
  complicated by the presence of deuterium, which will produce a
  Gunn-Peterson trough across the Ly$-n$ profiles for $x<81.5{\rm
    km\,s^{-1}}/b$. For a $D/H$ ratio of $2.8\pm0.2\times10^{-5}$
  \citep{2006ApJ...649L..61O}, the deuterium optical depths are
  $\tau_\alpha\simeq16.6\pm0.5$, $\tau_\beta\simeq2.6\pm0.1$ and
  $\tau_\gamma\simeq0.92\pm0.03$ at $z=8$. The IGM becomes optically
  thin to deuterium Ly-$n$ photons for $n>3$ at $z=8$ ($n>5$ at
  $z=20$). Since the results discussed in this paper rely
  predominantly on higher orders, as the direct collisional rates of
  these lower order Ly-$n$ photons (above \Lya) are already strongly
  suppressed, the effect of deuterium is not included.} In the Lorentz
wing, expressed as a function of the normalised frequency offset
$x=(\nu-\nu_{lu})/\Delta\nu_D$, the dimensionless Voigt profile
$\phi_V(a_n,x)=(\Delta\nu_D)\varphi_V(a_n,\nu)$ is well-approximated
by $\phi_V(a_n,x)\simeq a_n/(\pi x^2)$. The differential proper line
element evolves according to
$dl_p/dz=c/[H(z)(1+z)]\simeq(c/H_0)\Omega_m^{-1/2} (1+z)^{-5/2}$ in a
flat universe at redshifts for which $\Omega_m(1+z)^3$ dominates the
contribution from the vacuum energy, where $H(z)$ is the Hubble
parameter at redshift $z$. In the limit of scattering in the blue
wing, the optical depth along the path of a photon emitted at
frequency $x_e$ from a source at redshift $z_s$ and received at
frequency $x$ (provided it has not passed through any resonance line
centre en route) is given at large separations by
\begin{equation}
\tau_x\simeq x_1\left(\frac{1}{x}-\frac{1}{x_e}\right)
\qquad{\rm where}\qquad x_1=\frac{\sigma_n a_n \lambda_{lu} n_l(z)}
{\pi H(z)}
\label{eq:tauxi}
\end{equation}
\citep{2006MNRAS.372.1093F, 2009MNRAS.393..949H}, where $\lambda_{lu}$
is the wavelength of the resonance transition.

In terms of the atomic constants, $x_1$ may be expressed as
\begin{equation}
x_1=\frac{2}{\pi}\left(\frac{\pi e^2}{m_e
    c}\right)^2\frac{g_l}{g_u}f_{lu}^2\frac{n_l(z)}{H(z)b},
\label{eq:x1}
\end{equation}
where $m_e$ and $e$ are the mass and charge of the electron. For large
$n$ transitions, $f_{lu}\simeq (2^8/3 e^4) n^{-3}\simeq 1.56n^{-3}$,
so that $x_1$ decreases rapidly with increasing order as $x_1\sim
n^{-6}$. The effective mean free path of the photon as defined through
$\tau_x=l_p/l_{\rm mfp}$, where $l_p(z_S,z)=\int_{z}^{z_S}\,dz^\prime
dl_p/dz^\prime$ is the proper distance travelled by the photon, is
given by
\begin{equation}
  l_{\rm mfp}=\frac{\pi}{3}\left(\frac{\pi e^2}{m_e
      c}\right)^{-2}\frac{g_u}{g_l}f_{lu}^{-2}\frac{bc}{n_l(z)}
  \left[\left(1-\frac{1+z}{1+z_S}\right)^{3/2}\right]x,
\label{eq:mfp}
\end{equation}
so that the resonance line photon mean free path increases like $n^6$
for large $n$. For sufficiently large $n$, the mean free path will
match the proper distance $l_p$ for even small values of $x\sim{\cal
  O}(1)$. Thus, whilst the scattering of lower order Lyman resonance
line photons occurs in the Lorentz wing, at sufficiently high orders
the scattering will shift to the line core. The scattering of higher
order photons, however, is confined increasingly to the vicinity of
the source with increasing order. This is because, in an expanding
IGM, higher order Ly-$n$ photons may redshift into the next lower
order. The redshift $z_n$ to which a Ly-$n$ photon may travel from a
source at redshift $z_S$ is given by
$1+z_n=(1+z_S)[1-(n-1)^{-2}]/[1-n^{-2}]$.

\begin{figure}
\includegraphics[width=3.3in]{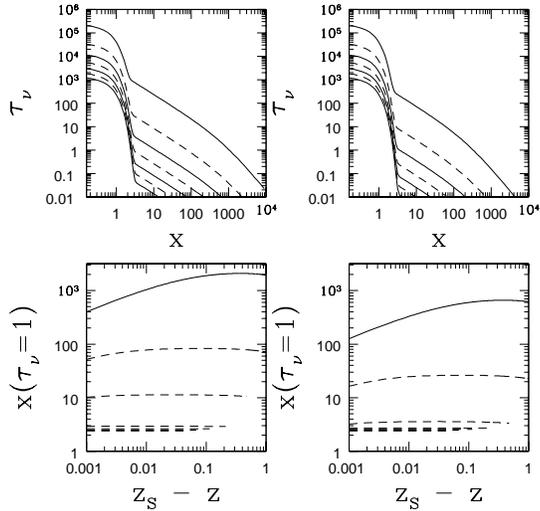}
\caption{(Upper panels)\ The optical depth of photons emitted by a
source at redshift $z_S=8$ and received at redshift $z=7.99$ at the
frequency $\nu$ near the
resonance line frequency $\nu_0$ for Ly$\alpha$ through Ly$\eta$ (top
to bottom, alternating solid and dashed curves). The
frequency is expressed as $x=(\nu-\nu_0)/\Delta\nu_D$, appropriate to
each order. The left panel shows the optical depth for an IGM
temperature of $T=10$~K and the right panel for $T=100$~K.
(Lower panels)\ The values $x_1$ of $x$ at which $\tau_\nu=1$ for
$0.001 < z_S - z < z_S-z_n$ for a source at redshift $z_S=8$, where
$z_n$ is the limiting redshift at which the frequency of an emitted
Ly-$n$ photon redshifts into the next lower Lyman resonance frequency.
The lines correspond, from top to bottom, to Ly$\alpha$ (solid line)
through Ly$\eta$ (dashed lines for higher Lyman orders).
The left panel shows the values of $x_1$ for an IGM
temperature of $T=10$~K and the right panel for $T=100$~K. (Not
included is the effect of deuterium; see text.)
}
\label{fig:taunu}
\end{figure}

\begin{table}
  \caption{Probabilities $p_{n,n^\prime}$ for a Ly-$n$ photon
    to convert into a Ly-$n^\prime$ photon per scatter. The final
    column gives the survival probability $p_{{\rm surv},
      n}=p_{n,n}$ for a Ly-$n$ photon to survive a scattering event
    as a Ly-$n$ photon.
  }
\begin{center}
\begin{tabular}{|c|l|l|l|l|l|} \hline
\hline 
$n$ & $p_{n,n^\prime=2}$ & $p_{n,n^\prime=3}$ &
 $p_{n,n^\prime=4}$ & $p_{n,n^\prime=5}$ & $p_{{\rm surv}, n}$\\

\hline
2  &1.000000 & 0.000000  & 0.000000 & 0.000000   & 1.000000  \\
3  &0.000000 & 0.881665  & 0.000000 & 0.000000   & 0.881665  \\
4  &0.041993 & 0.000000  & 0.839041 & 0.000000   & 0.839041  \\
5  &0.056103 & 0.007435  & 0.000000 & 0.817749   & 0.817749  \\
6  &0.062869 & 0.011134  & 0.002231 & 0.000000   & 0.805282  \\
7  &0.066774 & 0.013232  & 0.003553 & 0.000885   & 0.797245  \\
8  &0.069291 & 0.014555  & 0.004380 & 0.001464   & 0.791712  \\
9  &0.071035 & 0.015452  & 0.004935 & 0.001850   & 0.787718  \\
10 &0.072306 & 0.016094  & 0.005327 & 0.002120   & 0.784728  \\
11 &0.073269 & 0.016573  & 0.005617 & 0.002318   & 0.782423  \\
12 &0.074020 & 0.016941  & 0.005838 & 0.002468   & 0.780605  \\
13 &0.074619 & 0.017231  & 0.006011 & 0.002584   & 0.779142  \\ 
14 &0.075107 & 0.017465  & 0.006149 & 0.002677   & 0.777946  \\  
15 &0.075510 & 0.017657  & 0.006261 & 0.002751   & 0.776953  \\  
16 &0.075847 & 0.017816  & 0.006354 & 0.002813   & 0.776120  \\  
17 &0.076133 & 0.017950  & 0.006432 & 0.002865   & 0.775414  \\  
18 &0.076378 & 0.018064  & 0.006499 & 0.002908   & 0.774808  \\ 
19 &0.076590 & 0.018162  & 0.006555 & 0.002945   & 0.774285  \\
20 &0.076774 & 0.018247  & 0.006604 & 0.002977   & 0.773830  \\
21 &0.076936 & 0.018321  & 0.006646 & 0.003005   & 0.773432  \\
22 &0.077078 & 0.018386  & 0.006684 & 0.003029   & 0.773080  \\
23 &0.077205 & 0.018443  & 0.006716 & 0.003050   & 0.772769  \\
24 &0.077317 & 0.018494  & 0.006746 & 0.003069   & 0.772491  \\
25 &0.077419 & 0.018540  & 0.006771 & 0.003086   & 0.772243  \\
26 &0.077510 & 0.018581  & 0.006795 & 0.003101   & 0.772020  \\
27 &0.077592 & 0.018617  & 0.006816 & 0.003115   & 0.771818  \\
28 &0.077666 & 0.018651  & 0.006834 & 0.003127   & 0.771635  \\
29 &0.077734 & 0.018681  & 0.006852 & 0.003138   & 0.771469  \\
30 &0.077796 & 0.018709  & 0.006867 & 0.003148   & 0.771318  \\
31 &0.077853 & 0.018734  & 0.006881 & 0.003157   & 0.771179  \\
\hline
\end{tabular}
\end{center}
\label{tab:Lyn-to-Lynp}
\end{table}

The Lyman resonance order at which the transition to core scattering
occurs may be estimated as follows. The Voigt line profile is
well-approximated in the core by
$\phi_V(x)\simeq\pi^{-1/2}\exp(-x^2)$. For hydrogen gas in the
temperature range $10<T<1000$~K, the transition frequency $x_m$ at
which the core and wing approximations match lies in the range
$2.6<x_m<3.3$ for Lyman series photons with $n<8$. A second order
perturbation about $x_m=2.8$ gives the convenient approximation
$x_m\simeq2.8(1+\epsilon)$, with
$\epsilon\simeq[0.5987-0.1131(5.208+\log a)]^{1/2}-0.7738$ (accurate
to better than 0.1 per cent. for $0.001<a_n<0.05$). It follows from
Eq.~(\ref{eq:x1}) that Lyman resonance line photons with $n\gsim6$,
depending on the IGM temperature, will survive until redshifted into
the resonance line core. In Figure \ref{fig:taunu}, the intergalactic
optical depths through an IGM at temperatures $T=10$~K and 100~K are
shown for resonance line photons with $n=2-8$ emitted by a source at
$z_S=8$, along with the corresponding values of $x_1$ as a function of
redshift difference from the source. It is apparent that photons with
$n\ge5$ or 6 will first scatter in the resonance line core.

The source function describing the injection of Lyman series photons
following the initial scattering of continuum photons by the IGM is given by
\begin{eqnarray}
  S_n^{\rm inj}(\nu)&=&n_lc\sigma_n \int_0^\infty
  d\nu^\prime\,R_n(\nu^\prime,\nu)
  n_{\nu^\prime}^{\rm inc, n}\\
  &\simeq& n_l\sigma_n \frac{L_{\nu_n}}{4\pi r^2_L
    h_P\nu_n}\int_0^\infty
  d\nu^\prime\,R_n(\nu^\prime,\nu)\exp(-\tau_{\nu^\prime}),\nonumber
\label{eq:sourcefunction}
\end{eqnarray}
\citep{2009MNRAS.393..949H},where $n_\nu^{\rm inc, n}$ is the specific
number density of incident continuum photons near the resonance line
frequency $\nu_n$. The incident photon number density near Ly-$n$ is
given by $n_\nu^{\rm inc,n}=(4\pi/c)L_\nu \exp(-\tau_\nu)/[(4\pi
r_L)^2 h_P\nu]$ for a source of specific luminosity $L_\nu$ and at a
luminosity distance from the source $r_L$. It is clear that the source
functions of Ly-$n$ photons depend on the order, particularly as a
much smaller fraction of high order photons are scattered out of the
line of sight from the source than low order. As a consequence, the
scattering rate of higher order Lyman photons exceeds the rate of
lower order Lyman photons (above \Lya) for a nearly flat spectrum
source.

The case of \Lya\ photons is special, since these photons cannot
degrade. Instead, because the IGM is optically thick to \Lya\ photons
the photons become trapped in the IGM until they redshift through the
\Lya\ resonance as a consequence of cosmological expansion. A balance
is then achieved between the injection rate and the escape rate of the
photons from the IGM. The number of scatters ${\cal N}_{\rm scat}$
before the photons escape is given by the inverse of the Sobolev
expansion parameter $\gamma$, which is the inverse of the \Lya\
optical depth in the expanding IGM, ${\cal N}_{\rm
  scat}=\gamma^{-1}=\tau_\alpha=\lambda_\alpha\sigma_\alpha
n_l(z)/H(z)$ \citep{1959ApJ...129..536F, 1965ApJ...142.1633G,
  MMR97}.\footnote{The equivalence ${\cal N}_{\rm scat}=\gamma^{-1}$
  also follows from the IGM scattering problem of
  \citet{2009MNRAS.393..949H}, who obtain an equilibrium photon number
  density $n_\nu=\dot s/(n_lc\sigma_\alpha\gamma)$ in the red wing
  produced by the redshifting of photons following their injection by
  a Dirac $\delta$-function source of strength $\dot s$ at a frequency
  $\nu_{\rm inj}$, describing the trapping and rescattering of photons
  in the blue wing by the IGM. In equilibrium, the photons in the red
  wing must redshift away at the rate $\int d\nu\,n_\nu/({\cal N}_{\rm
    scat}t_{\rm scat})=\dot s$, giving ${\cal N}_{\rm
    scat}=\gamma^{-1}$.} The rate of photon scatters per hydrogen
atom, given by integrating Eq.~(5) over frequency, is then
$n_l^{-1}\int\,d\nu S_\alpha^{\rm inj}(\nu)=n^{\rm
  inc,\alpha}_\nu(0)c\sigma_\alpha (a_\alpha/\pi) x_1^{-1}$,
approximating the Voigt profile by $(a_\alpha/\pi)x^{-2}$. Multiplying
by the number of scatters ${\cal N}_{\rm scat}$ gives for the
scattering rate per hydrogen atom, $P_\alpha^{\rm direct}=n^{\rm inc,
  \alpha}_\nu(0)c\sigma_\alpha$ since
$(a_\alpha/\pi)(\tau_\alpha/x_1)=1$. Thus the scattering rate is
simply given by the scattering rate of the incident photons assuming
no scattering losses en route \citep{1959ApJ...129..536F, MMR97,
  2009MNRAS.393..949H}.

By contrast, Lyman resonance line photons above \Lya\ that first
scatter in the wings will produce a negligible scattering rate because
they will quickly degrade into a lower energy photon before they can
redshift or diffuse into the resonance line core. A typical photon
will not survive more than about five scatters
\citep{2006MNRAS.367.1057P}. As will now be shown, higher order Lyman
series photons that first scatter in the resonance line core will
produce a non-negligible scattering rate, contrary to previous
estimates which assumed a constant source function for all the Lyman
resonance line photons \citep{2005ApJ...626....1B,
  2006MNRAS.367..259H, 2006MNRAS.367.1057P}, leading to a scattering
rate estimate of order ${\cal O}(5\times10^{-6})$ times smaller than
the \Lya\ rate.

The scattering rate per atom will depend on the total number density
of resonance line photons built up in the region in which the incident
radiation field generates them. For a mean scattering time of $t_{{\rm
    scat},n}(\nu)=1/n_lc\sigma_n\phi_V(a_n,\nu)$ and a photon survival
probability $p_{{\rm surv},n}$, describing the fraction of Ly-$n$
photons that produce a subsequent Ly-$n$ photon upon scattering, the
contribution of direct Ly-$n$ scatterings from the source to the
photon number density, including the subsequent rescatterings of the
photons, is given by
\begin{eqnarray}
  n^{{\rm direct}, n}_\nu&=&n^{{\rm inc}, n}_\nu\nonumber\\
  &+&n_lc\sigma_np_{{\rm surv}, n}t_{{\rm scat}, n}(\nu)\nonumber\\
&\times&\int_0^\infty d\nu^\prime\,R_n(\nu^\prime,\nu)n^{\rm inc,
    n}_{\nu^\prime}\nonumber\\
  &+&\dots\nonumber\\
  &+&(n_lc\sigma_np_{{\rm surv}, n})^{m+1}t_{{\rm scat}, n}(\nu)\nonumber\\
&\times&\int_0^\infty d\nu_m\, R_n(\nu_m,\nu)t_{{\rm scat}, n}(\nu_m)\dots\nonumber\\
  &\times&\int_0^\infty d\nu_1\, R_n(\nu_1,\nu_2)t_{{\rm scat}, n}(\nu_1)\nonumber\\
&\times&\int_0^\infty d\nu^\prime\, R_n(\nu^\prime,\nu_1)n^{{\rm inc},
  n}_{\nu^\prime}\nonumber\\
&+&\dots .
\label{eq:ndirect}
\end{eqnarray}
Here, $R_n(\nu^\prime,\nu)$ is the frequency redistribution function
for the scattered photons. The survival probabilites $p_{{\rm
    surv},n}$ follow from the spontaneous transition rates
$A(n,l;n^\prime,l^\prime)$ according to
\begin{equation}
  p_{{\rm surv}, n}=A_{n,1}^{-1}A(n,1;1,0),
\label{eq:psurv}
\end{equation}
where $A_{n,l}=\Sigma_{j=1}^{n-1}\Sigma_{k=0}^{j-1}A(n,l;j,k)$ denotes
the total decay rate from level $(n,l)$ to all lower levels. Values
for $n=2$ to 31 are provided in Table~\ref{tab:Lyn-to-Lynp}.

\begin{table}
  \caption{The ratio ${\cal N}_{\nu_n}/{\cal N}_{\nu_n}^{\rm inc}(0)$
    of Ly-$n$ photon occupation number to the occupation
    number from the incident radiation field assuming an optically
    thin IGM, divided by the IGM suppression
    factor ${\cal S}_n$, followed by the value of the suppression factor,
    for a source at redshift $z_S=8$ and a $T=10$~K IGM, at (proper)
    distances from the source of 20~kpc (cols 2--3) and 100~kpc (cols
    5--6). For \Lya, only ${\cal N}_\alpha/{\cal N}_\alpha^{\rm
      inc}(0)$ is shown. Also provided are the corresponding light
    temperatures $\langle T_n\rangle_{\rm H}$ due to the direct
    radiation field $n^{{\rm direct}, n}_\nu$ (cols 4 and 7).
  }
\begin{center}
  \begin{tabular}{|c|l|l|l|l|l|l|} \hline \hline $n$ & $\frac{{\cal
        N}_{\nu_n}}{{\cal N}_{\nu_n}^{\rm inc}(0) {\cal S}_n}$ & ${\cal
    S}_n$ & $\langle T_n\rangle_{\rm H}$ & $\frac{{\cal
      N}_{\nu_n}}{{\cal N}_{\nu_n}^{\rm inc}(0){\cal
      S}_n}$ & ${\cal S}_n$ & $\langle T_n\rangle_{\rm H}$ \\
  \hline
  2    &8.88360 & $\dots$  & $\dots$ & 5.11169 & $\dots$  & $\dots$ \\
  3    &17.4419 & 0.000022 & -6.1130 & 14.7196 & 0.000015 & -7.6446 \\
  4    &9.77899 & 0.000042 & -0.6605 & 8.02393 & 0.000037 & -0.6873 \\
  5    &7.14941 & 0.000075 & -0.0737 & 6.20837 & 0.000071 & -0.0739 \\
  6    &5.89017 & 0.000137 & -0.0483 & 5.42872 & 0.000135 & -0.0483 \\
  7    &5.32458 & 0.000225 & -0.0440 & 5.06954 & 0.000224 & -0.0440 \\
  8    &5.02326 & 0.000342 & -0.0436 & 4.87066 & 0.000342 & -0.0436 \\
  9    &4.84420 & 0.000494 & -0.0442 & 4.74756 & 0.000494 & -0.0442 \\
  10   &4.72910 & 0.000684 & -0.0451 & 4.66522 & 0.000683 & -0.0451 \\
  11   &4.65052 & 0.000916 & -0.0462 & 4.60691 & 0.000916 & -0.0462 \\
  12   &4.59430 & 0.001195 & -0.0473 & 4.56379 & 0.001195 & -0.0473 \\
  13   &4.55252 & 0.001525 & -0.0483 & 4.53078 & 0.001526 & -0.0483 \\
  14   &4.52051 & 0.001910 & -0.0494 & 4.50481 & 0.001911 & -0.0494 \\
  15   &4.49536 & 0.002356 & -0.0505 & 4.48394 & 0.002356 & -0.0505 \\
  16   &4.47516 & 0.002864 & -0.0516 & 4.46684 & 0.002866 & -0.0516 \\
  17   &4.45866 & 0.003441 & -0.0526 & 4.45263 & 0.003443 & -0.0526 \\
  18   &4.44495 & 0.004091 & -0.0537 & 4.44066 & 0.004092 & -0.0537 \\
  19   &4.43343 & 0.004817 & -0.0548 &&\\
  20   &4.42363 & 0.005624 & -0.0558 &&\\
  21   &4.41521 & 0.006516 & -0.0569 &&\\
  22   &4.40791 & 0.007497 & -0.0579 &&\\
  23   &4.40153 & 0.008572 & -0.0590 &&\\
  24   &4.39592 & 0.009745 & -0.0601 &&\\
  25   &4.39095 & 0.011020 & -0.0612 &&\\
  26   &4.38652 & 0.012402 & -0.0623 &&\\
  27   &4.38256 & 0.013894 & -0.0634 &&\\
  28   &4.37900 & 0.015501 & -0.0645 &&\\
  29   &4.37579 & 0.017227 & -0.0657 &&\\
  30   &4.37288 & 0.019077 & -0.0668 &&\\
  31   &4.37024 & 0.021055 & -0.0680 &&\\
  \hline
\end{tabular}
\end{center}
\label{tab:supn}
\end{table}

The direct Ly-$n$ scattering rate is given by integrating $n^{{\rm
  direct}, n}_\nu$ over the Ly-$n$ cross section. Noting that $\int
d\nu R_n(\nu^\prime,\nu)=\varphi_V(a_n,\nu^\prime)$, the integrals in
Eq.~(\ref{eq:ndirect}) contract, resulting in the direct scattering rate
\begin{eqnarray}
  P^{\rm direct}_n &=& \int_0^\infty d\nu\, n_{\rm \nu}c\sigma_n
\varphi_V(a_n,\nu)\nonumber\\
  &=&P^{\rm inc}_n/(1-p_{{\rm surv}, n}),\nonumber\\
  &=&P^{\rm inc}_n(0){\cal S}_n/(1-p_{{\rm surv}, n}),
\label{eq:Pndirect}
\end{eqnarray}
where $P_n^{\rm inc}(0)=n^{{\rm inc},n}_\nu(0)c\sigma_n= \sigma_n
L_{\nu_n}/(4\pi r_L^2 h_P\nu_n)$ is the scattering rate for an
optically thin IGM, and
\begin{equation}
{\cal S}_n=\int_{0}^\infty
  d\nu^\prime\,\varphi_V(a_n,\nu^\prime)\exp(-\tau_{\nu^\prime})
\label{eq:Sn}
\end{equation}
describes the scattering suppression factor due to the scattering out
of source photons by the intervening IGM out to the distance $r_L$
\citep{2008MNRAS.390.1430D,2009MNRAS.393..949H}. Values of ${\cal
  S}_n$ are tabulated in Table~\ref{tab:supn} for some typical
situations.

Although the expression in Eq.~(\ref{eq:Pndirect}) follows from a
seemingly straightforward account of photon scatters, the expression
for $n^{\rm direct, n}_\nu$ in Eq.~(\ref{eq:ndirect}) reveals that an
implicit non-trivial assumption has been made:\ it has been assumed
that the photons rescattered from frequency $\nu_j$ to $\nu_{j+1}$
have sufficient time to do so before their frequencies are Doppler
shifted relative to the expanding (or contracting gas). Since the
photons are initially scattered blueward of the line centre,
redshifting will generally carry the photons to frequencies with
larger scattering optical depths. If the redshifting timescale
(characterised by the inverse of the divergence of the velocity
field), is shorter than the scattering time at the frequency $\nu_j$,
the redshifting timescale should be used in Eq.~(\ref{eq:ndirect}). In
the case of a contracting region, blueshifting may carry the photons
to regions of a longer mean free path, and hence longer scattering
time. If the scattering time exceeds the contraction timescale, then
scattering may not occur at all before the photons are blueshifted
away, and so the contraction timescale should be used. These factors
are further complicated by the redistribution of the photon
frequencies by the scattering, which tends to drive photons towards
the line centre for scatters not too far out in the wing, and to
concentrate the photon frequencies at the incoming frequency for
scatters well into the wings (eg, \citet{1978stat.book.....M}). In the
event the scattering time is shorter than the characteristic expansion
or contraction timescale, Eq.~(\ref{eq:ndirect}) may be used, from
which Eq.(\ref{eq:Pndirect}) follows.

Incident resonance line Ly-$n$ photons that do not rescatter as Ly-$n$
photons will produce lower order Lyman series photons through the
ensuing radiative cascades, adding to the direct scattering rate
above. The cascade rates may be estimated similarly to the above,
except that the redistribution function must now describe the
conversion of a Ly-$n$ photon to Ly-$n^\prime$ ($n^\prime<n$). The
Lyman photon conversion probabilities per scatter $p_{n,n^\prime}$ may
be computed from the probability for an atom in state $(n,l)$ to
produce a Ly-$n^\prime$ photon through radiative cascades,
\begin{equation}
  p_{\rm casc}(n,l;n^\prime)=A_{n,l}^{-1}\sum _{j=n^\prime}^{n-1}
  \sum_{k=0}^{j-1}A(n,l;j,k) p_{\rm casc}(j,k;n^\prime).
\label{eq:pnnp}
\end{equation}
The values for $p_{\rm casc}(j,k;n^\prime)$ may be computed
iteratively from low $j$ to high, initiated by $p_{\rm
  casc}(n^\prime,1;n^\prime)=p_{{\rm surv},n^\prime}$ and $p_{\rm
  casc}(n^\prime,k;n^\prime)=0$ for $k\ne1$. The Lyman photon
conversion probabilities are then given by $p_{n,n^\prime}=p_{\rm
  casc}(n,1;n^\prime)$. Values for $n=2$ to 31 and $n^\prime\leq5$ are
provided in Table~\ref{tab:Lyn-to-Lynp}.

The photon density $n^{{\rm direct}, n}_\nu$ serves to source lower
order photons through scatterings in which the original Ly-$n$ photon
degrades into lower energy photons. Since every order will produce
further lower order Lyman photons through scatterings, a production
cascade of Lyman photons results. The rates may be computed by
starting with the highest order Lyman resonance line photons, $n_{\rm
  max}$, reaching a given distance from the central source. The
production rate of the next lower allowed Lyman order photons is
\begin{eqnarray}
  S_{n_{\rm max}, n_{{\rm max}-2}}(\nu)&=&n_l\sigma_{n_{\rm max}}p_{n_{\rm
      max},{n_{\rm max}-2}}\\
  &\times&\int_0^\infty d\nu^\prime R_{n_{\rm
      max},n{{\rm max}-2}}(\nu^\prime,\nu)n^{{\rm direct}, n_{\rm
      max}}_{\nu^\prime},\nonumber
\label{eq:Snmaxnmaxm2}
\end{eqnarray}
where $R_{n^\prime,n}(\nu^\prime,\nu)$ is the redistribution function
describing the scattering of a Ly-$n^\prime$ photon at frequency
$\nu^\prime$ to a Ly-$n$ photon at frequency $\nu$. The redistribution
function is discussed in the Appendix. The resulting density of
cascade-produced $n_{{\rm max}-2}(\nu)$ photons, including their
rescatterings, is then given by
\begin{eqnarray}
  n^{{\rm cascade}, {n_{\rm max}-2}}_\nu&=&n_lc\sigma_{n_{\rm max}}
p_{n_{\rm max}, n_{{\rm max}-2}}t_{{\rm scat}, n_{\rm max}-2}(\nu)\nonumber\\
&\times&\Biggl[\int_0^\infty d\nu^\prime\,
R_{n_{\rm max},n_{{\rm max}-2}}(\nu^\prime,\nu)n^{{\rm direct},
    n_{\rm max}}_{\nu^\prime}\nonumber\\
&+&n_lc\sigma_{n_{\rm max}-2}p_{{\rm
    surv}, n_{{\rm max}-2}}\nonumber\\
&\times&\int_0^\infty d\nu_1\, R_{n_{\rm max}-2}(\nu_1,\nu)t_{{\rm
    scat}, n_{\rm max}-2}(\nu_1)\nonumber\\
  &\times&\int_0^\infty d\nu^\prime\, R_{n_{\rm max},n_{\rm
      max}-2}(\nu^\prime,\nu_1)n^{{\rm direct},
    n_{\rm max}}_{\nu^\prime}\nonumber\\
  &+&\dots\nonumber\\
  &+&(n_lc\sigma_{n_{\rm
      max}-2}p_{{\rm surv}, n_{{\rm max}-2}})^m\nonumber\\
&\times&\int_0^\infty d\nu_m\, R_{n_{\rm max}-2}(\nu_m,\nu)
t_{{\rm scat}, n_{{\rm max}-2}}(\nu_m)\dots\nonumber\\
  &\times&\int_0^\infty d\nu_1\, R_{n_{\rm max}-2}(\nu_1,\nu_2)
t_{{\rm scat}, n_{\rm max}-2}(\nu_1)\nonumber\\
&\times&\int_0^\infty d\nu^\prime\, R_{n_{\rm max},n_{\rm
    max}-2}(\nu^\prime,\nu_1)n^{{\rm direct},
    n_{\rm max}}_{\nu^\prime}\nonumber\\
&+&\phantom{\Biggl[}\dots\Biggr].
\label{eq:ncascade}
\end{eqnarray}

As in the case of $n^{{\rm direct},n}_\nu$ in Eq.~(\ref{eq:ndirect}),
it has been implicitly assumed that the photons are not much redshifted (or
blueshifted) relative to the scattering medium if it is expanding (or
contracting) before being scattered. If not, then integrating $n^{{\rm
    cascade},n_{\rm max}-2}_\nu$ over the scattering cross-section contracts
Eq.~(\ref{eq:ncascade}) into
\begin{equation}
P^{\rm direct-cascade}_{n_{\rm max}, n_{\rm max}-2}=\frac{p_{n_{\rm
      max}, n_{\rm max}-2}}{1-p_{{\rm surv},n_{\rm max}-2}}P^{\rm
  direct}_{n_{\rm max}}.
\label{eq:Pdircasnmax}
\end{equation}
A similar expression results for the scattering of any directly generated
Ly-$n^\prime$ photon into a Ly-$n$ photon,
\begin{equation}
P^{\rm direct-cascade}_{n^\prime, n}=\frac{p_{n^\prime, n}}{1-p_{{\rm
      surv},n}}
P^{\rm  direct}_{n^\prime}.
\label{eq:Pndircasnp}
\end{equation}

Each cascade generated Ly-$n$ photon in turn serves as a source of
further lower order Lyman photons. Considerations similar to the above
result in the net cascade-generated collision rate
\begin{equation}
P^{\rm cascade}_n=\frac{1}{1-p_{{\rm surv},n}}
\sum_{n^\prime=n+1}^{n_{\rm max}}p_{n^\prime,n}P_{n^\prime},
\label{eq:Pncascade}
\end{equation}
where the total scattering rate of Ly-$n^\prime$ photons is
$P_{n^\prime}=P^{\rm direct}_{n^\prime}+P^{\rm
  cascade}_{n^\prime}$. The cascade scattering rates may be solved for
iteratively starting from higher order to low with $P_{n_{\rm
    max}}=P^{\rm direct}_{n_{\rm max}}=P^{\rm inc}_{n_{\rm
    max}}/(1-p_{{\rm surv},n_{\rm max}})$. For \Lya\ photons ($n=2$),
the same expression may be used except that $1/(1-p_{{\rm surv},n})$
must be replaced by $N_{\rm scat}=\tau_\alpha$ to account for the
accumulation of \Lya\ photons as they grow in density once trapped in
the IGM, ultimately redshifting away. It is the set of total rates
$P_n$ that drives collisional photon heating and the W-F effect
extended to higher Lyman orders.

It will be convenient below to express the scattering rates in terms
of the mean photon occupation number of the resonance line photons
averaged over the resonance line scattering profile ${\cal
  N}_{\nu_n}=(c/8\pi)\lambda_n^2\int_0^\infty d\nu\,
n_\nu\phi_V(a_n,\nu)$, where $n_\nu$ is the specific photon number
density including the enhancements due to cascades from any higher
order Lyman series photon scatterings and rescatterings. The
scattering rate per atom in the lower state is then
$P_n=(g_u/g_l)A_{ul}{\cal N}_{\nu_n}$, where $g_u$ and $g_l$ are the
statistical weights of the upper and lower states, respectively, and
$A_{ul}$ is the spontaneous decay rate for the transition. In terms of
the occupation numbers, Eq.~(\ref{eq:Pndirect}) and
Eq.~(\ref{eq:Pncascade}) may be used to express the contributions from
direct scatters and subsequent cascades to the photon occupation
numbers (for $n>2$) as
\begin{equation}
{\cal N}_{\nu_n}^{\rm direct}={\cal N}_{\nu_n}^{\rm inc}(0){\cal S}_n/(1-p_{{\rm
    surv},n}),
\label{eq:Noccndirect}
\end{equation}
and
\begin{equation}
{\cal N}_{\nu_n}^{\rm cascade}=\frac{1}{1-p_{{\rm
      surv},n}}\sum_{n^\prime=n+1}^{n_{\rm max}}
\frac{A(n^\prime,1;1,0)}{A(n,1;1,0)}p_{n^\prime,n}{\cal N}_{\nu_{n^\prime}},
\label{eq:Noccncascade}
\end{equation}
where ${\cal N}_{\nu_{n^\prime}}={\cal N}_{\nu_{n^\prime}}^{\rm direct}+
{\cal N}_{\nu_{n^\prime}}^{\rm cascade}$ is the total occupation number of
Ly-$n^\prime$ photons. For \Lya\ photons,
\begin{equation}
{\cal N}_\alpha^{\rm cascade}=N_{\rm scat}\sum_{n=4}^{n_{\rm max}}
\frac{A(n,1;1,0)}{A(2,1;1,0)}p_{n,2}{\cal N}_{\nu_n}.
\label{eq:Noccacascade}
\end{equation}
The mean photon occupation number for the Ly-$n$
transition will also sometimes be indicated by ${\cal N}_{lu}$ where
greater specification of the energy levels involved is required. In
particular, for \Lya\ photons in statistical equilibrium with the gas,
${\cal N}_{si}={\cal N}_{ti}\exp(-T_{st}/T_L)$, where ${\cal N}_{si}$
and ${\cal N}_{ti}$ describe the \Lya\ photon occupation numbers
corresponding to the hyperfine $n=2$ states $i$ and the hyperfine
ground state singlet and triplet states, respectively. For higher
levels ($n>2$), ${\cal N}_{si}={\cal N}_{ti}$ may be taken.

The spontaneous decay rates $A(n,l;n^\prime,l^\prime)$ are computed
following \citet{1963tas..book.....C}. Excellent agreement is found
with the rates published in \citet{1966atp..book.....W}. The survival
probabilities of Ly-$n$ photons computed agree precisely with those
tabulated by \citet{2006MNRAS.367.1057P}, and the \Lya\ production
probabilities $p_{n,2}$ agree precisely with those tabulated by
\citet{2006MNRAS.367..259H} and \citet{2006MNRAS.367.1057P}, who cite
the fraction of degraded Ly-$n$ photons that produce \Lya\ photons,
expressed in terms of the values given here by $p_{n,2}/(1-p_{{\rm
    surv},2})$.

Representative photon occupation numbers including rescatterings and
cascades are shown in Table~\ref{tab:supn}. The values are normalised
by the occupation number of incident Ly-$n$ photons assuming an
optically thin IGM. A source with constant $L_\lambda$ is assumed, as
this approximates a starburst spectrum at wavelengths just longward of
the Lyman edge \citep{1999ApJS..123....3L}. Whilst rescatterings
increase the numbers of higher order Ly-$n$ photons by the factor
$1/(1-p_{{\rm surv}, n})$ above the incident number, cascades add
little more. The exception is for \Lya\ photons, for which cascades
substantially boost the photon density because the \Lya\ photons
become trapped in the IGM, leaving only as they redshift through the
resonance line frequency. Approximately one third more \Lya\ photons
are obtained compared with previous findings
\citep{2005ApJ...626....1B, 2006MNRAS.367..259H, 2006MNRAS.367.1057P,
  2007ApJ...670..912C}. Instead of Eq.~(\ref{eq:Noccacascade}), these
earlier estimates used the equivalent of ${\cal N}_\alpha^{\rm
  cascade}=\sum_{n=4}^{n_{\rm max}}p_{n,2}(\nu_n/\nu_\alpha)^2{\cal
  N}_{\nu_n}^{\rm inc}(0)/(1-p_{{\rm surv},n})$, where ${\cal
  N}_{\nu_n}^{\rm inc}(0)$ is the occupation number of the incident
Ly-$n$ photons in an optically thin IGM. For a source at $z_S=8$ in a
$T=10$~K IGM, Eq.~(\ref{eq:Noccacascade}) gives 29 per cent. more
cascade-produced \Lya\ photons at a proper separation of 1~Mpc from
the source, 32 per cent. more at 100~kpc, and 33 per cent. more at
20~kpc.

It is emphasised that radiative transfer effects for \Lya\ photons
involving spatial diffusion, which may enhance the density of \Lya\
photons near a source \citep{1999ApJ...524..527L, 2007ApJ...670..912C,
  2007A&A...474..365S}, or recoils, which may suppress the density
near line center and so reduce the scattering rate
\citep{2004ApJ...602....1C, 2009MNRAS.393..949H} have not been
included. It is presumed that the \Lya\ photons produced in cascades
will be subject to the same radiative transfer processes as the
directly incident photons, so that the enhancements here reflect the
fractional increase due to cascades alone. This is not obviously
correct, particularly as the injected photons may arrive within the
line core, as opposed to the wings as assumed in most previous
studies. The actual degree of enhancement will depend on the relative
effects of spatial diffusion, recoil, the local rate of expansion or
contraction of the scattering medium, and escape from substructures
within the IGM.

\section{Collisional heating by Lyman resonance line photons}
\label{sec:heating}

When \Lya\ photons are first incident on cold neutral hydrogen gas,
they provide a source of heat resulting from the momentum transferred
to the atoms by the scattering photons \citep{MMR97}. This ``recoil
heating'' persists for only a short period before the radiation field
reaches statistical equilibrium with the gas through multiple
scatters, and establishes thermal equilibrium with the gas
\citep{1959ApJ...129..551F, MMR97, Meiksin06}. Thereafter at most a
residual amount of energy transfer remains in a cosmological setting,
scaling like $\gamma$ as photons redshift through the resonance line
frequency \citep{2004ApJ...602....1C}. For typical IGM conditions at
high redshifts, the time to achieve thermal equilibrium is about
1--10~yrs, corresponding to several tens to hundreds of scatters
\citep{Meiksin06}; \Lya\ photon collisional heating subsequently
becomes an inefficient heating mechanism.

By contrast, because higher order Lyman resonance line photons do not
survive long enough to establish thermal equilibrium with the
scattering medium before degrading into lower energy photons, they may
provide a significant source of heating provided their scattering rate
is sufficiently high. The analysis for Ly-$n$ photons that rescatter
as Ly-$n$ photons is identical to that for \Lya\ photons. In addition
to order-preserving scatterings (Ly-$n$ to Ly-$n$), scatterings in
which the incident Lyman photon degrades into lower energy photons
(eg, Ly$\gamma$ into Pa$\alpha$, H$\alpha$ and \Lya), also provide a
source of recoil heating. In this case only the photon produced in the
scattering event provides a recoil to the atom, not the subsequent
decays, as may be demonstrated as follows.

The 4-momentum of an incoming atom of rest mass $m_a$ and quantum
state energy $\epsilon_i$ is $p_{ai}=[\gamma_i(m_a +
\epsilon_i/c^2){\bf v}_i,\gamma_i(m_ac+\epsilon_i/c)]$. The 4-momentum
of the incoming photon which it scatters is $p_{\gamma
  i}=(h\nu^\prime/c)({\bf {\hat n}^\prime},1)$. After the scattering
event, the 4-momenta of the atom and photon are
$p_{af}=[\gamma_f(m_a+\epsilon_f/c^2) {\bf v}_f,
\gamma_f(m_ac+\epsilon_f/c)]$ and $p_{\gamma f}=(h\nu/c)({\bf \hat
  n},1)$, for a final atomic quantum state of energy $\epsilon_f$. Here,
$\gamma_{i,f}=(1-v_{i,f}^2/c^2)^{-1/2}$. To lowest order in $v/c$, and
neglecting $\epsilon_i$ and $\epsilon_f$ relative to the rest mass
energy of the atom, the resulting frequency shift between the outgoing
and incoming photons becomes
\begin{eqnarray}
(\nu&-&\nu^\prime)\Biggl(1-\frac{\bf v_i}{c}\cdot{\bf\hat n}\Biggr)\nonumber\\
&\simeq&\nu^\prime\frac{\bf v_i}{c}\cdot({\bf\hat n} -
{\bf{\hat n}^\prime})-\frac{h\nu\nu^\prime}{m_ac^2}(1-{\bf\hat
n}\cdot{\bf{\hat n}^\prime}) - \frac{\epsilon_f-\epsilon_i}{h}.
\label{eq:Deltanu}
\end{eqnarray}
The expression reflects the change in frequency between the incoming
and outgoing photons, including the change in the energy state of the
atom, the Doppler shift due to the motion of the atom, and the recoil,
which depends on the ratio $h\nu\nu^\prime/m_ac^2$ and corresponds on
average to an energy loss from the radiation field. This latter terms
gives rise to the heating of the IGM.

The frequency of an outgoing photon following the spontaneous decay of
the atom after a scattering event may be expressed as the limit of an
incoming photon with $\nu^\prime=0$. The frequency of the outgoing
photon is then
\begin{equation}
\nu=\frac{\epsilon_i-\epsilon_f}{h(1-{\bf v_i}\cdot{\bf\hat n}/c)},
\label{eq:Deltanudecay}
\end{equation}
which expresses the energy difference of the atom and the Doppler
shift due to its motion, without any recoil term. As a consequence,
whilst recoil due to the scattering of a photon which degrades upon
scattering must be accounted for, no recoils result from any
subsequent decays of the atom in the non-relativistic limit.

It is shown in the Appendix that the contribution from scatterings in
which the incoming photon is degraded is identical in form to that of
a surviving photon:\ it is only the absorption line profile and the
incident and scattered photon energies that enter into the average
amount of energy exchanged with the gas. (Any subsequent Ly-$n$
photons produced following scattering events in which the original
photon is degraded, of course, will also contribute to the recoil
heating term if scattered, at a rate depending on the corresponding
absorption line profile.)

\begin{figure}
\includegraphics[width=3.3in]{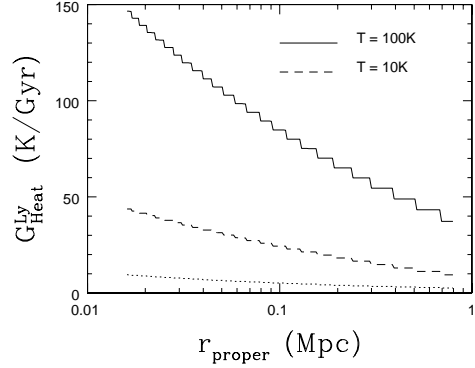}
\caption{Lyman photon collisional heating rate for a source at $z_S=8$
  and IGM temperatures $T=100$~K (solid line) and $T=10$~K (dashed
  line), as a function of proper distance from the source. The heating
  rate is normalised to a \Lya\ scattering rate equal to the
  thermalisation rate $P_{\rm th}\simeq6.8\times10^{-12}[(1+z)/9]\,
  {\rm s^{-1}}$ required to bring the spin temperature to the kinetic
  temperature of the IGM through the W-F mechanism. Also shown (dotted
  line) is the total \Lya\ scattering rate $P_\alpha/P_\alpha^{\rm
    inc}(0)$ normalised by the scattering rate of the incident \Lya\
  photons not including the contribution from cascades from higher
  Lyman orders.
}
\label{fig:GHeat}
\end{figure}

The resulting total heating rate due to the scattering of Ly-$n$
photons is
\begin{eqnarray}
G_n = P_n n_l \frac{h\nu_n}{m_a c^2}&\sum_{n^\prime=1}^{n-1}&
\frac{A(n,1;n^\prime,0) + A(n,1;n^\prime,2)}{A_{n,1}}\nonumber\\
&\times&h\nu_{nn^\prime}
\Biggl(1-\frac{T}{\langle T_{nn^\prime}\rangle_{\rm H}}\Biggr),
\label{eq:Gn}
\end{eqnarray}
for an IGM of temperature $T$ and a harmonic mean light temperature of
\begin{equation}
\langle T_{nn^\prime}\rangle_{\rm H}=\frac{\nu_{nn^\prime}}{\nu_n}
\frac{\int_0^\infty d\nu\, n_\nu\varphi_V(a_n,\nu)}{\int_0^\infty d\nu\,
\frac{1}{T_n(\nu)}n_\nu\varphi_V(a_n,\nu)},
\label{eq:TnH}
\end{equation}
where
\begin{equation}
T_n(\nu)=-\frac{h}{k_{\rm B}}\Biggl(\frac{d\log
  n_\nu}{d\nu}\Biggr)^{-1}
\label{eq:Tnnu}
\end{equation}
(cf \citet{Meiksin06}). Here
$\nu_{nn^\prime}=\nu_L(1/{n^\prime}^2-1/n^2)$ (and $\nu_n=\nu_{n1}$),
where $\nu_L$ is the frequency of the Lyman edge. $A_{n,1}$ is the
total decay rate of the $p$-state with principal quantum number
$n$. (Note $A(n,1;n^\prime,2)$ is undefined for $n^\prime<3$, and
should be regarded as zero.)

The frequency distribution of the photons is determined both by the
incident radiation field and the photons produced through
rescatterings and cascades. It is convenient to consider the
contributions from these secondary photons separately.

Computing the light temperature for the secondary photons requires
integrations over the redistribution functions. An estimate may be
made most simply for Ly-$n$ photons rescattered into Ly-$n$. For
scatters in the blue wing, the frequency redistribution may be
approximated by coherent scattering for which
$R_n(\nu^\prime,\nu)\simeq\varphi_V(a_n,\nu)\delta_D(\nu-\nu^\prime)$,
where $\delta_D$ is the Dirac
$\delta$-function. Eq.~(\ref{eq:ndirect}) then contracts to $n^{{\rm
    direct}, n}_\nu\simeq n^{{\rm inc}, n}_\nu/(1-p_{{\rm surv}, n}) =
n^{{\rm inc}, n}_\nu(0){\cal S}_n/(1-p_{{\rm surv}, n})$. For a
negligible contribution from the core, the light temperature is then
given by
\begin{eqnarray}
\langle T_n^{\rm wing}\rangle_{\rm H}&\simeq&-\frac{h\nu_L}{2k_{\rm
    B}}
\left(1-\frac{1}{n^2}\right)\frac{b}{c}x_1\nonumber\\
&\times&\frac{1-\exp(-x_1/x_m)}{1-(1+x_1/x_m)\exp(-x_1/x_m)}\nonumber\\
&\simeq&-0.0338\left(1-\frac{1}{n^2}\right)T^{1/2}x_1\,{\rm K},
\label{eq:TLwing}
\end{eqnarray}
where $x_1>>x_m$ has been assumed in the last line. The light
temperature is independent of the gas temperature for wing scatters
since $x_1\sim T^{-1/2}$. The skewness of the photon number density
towards the blue results in a negative light temperature,
corresponding to a net transfer of heat to the gas resulting from the
Doppler shifting of the photons by the gas. This is in addition to the
recoil heating term, and will contribute a comparable amount. Because
of the small value of the suppression factor ${\cal S}_n$, however,
the scattering rate will be small, and the heating rate as well.

A larger source of heating will arise from Ly-$n$ photons that first
scatter in the core $x_1\lta x_m$. In this case, the redistribution
function may be approximated in the limit of complete redistribution,
$R_n(\nu^\prime,\nu)\simeq\varphi_V(a_n,\nu^\prime)\varphi_V(a_n,\nu)$.
Eq.~(\ref{eq:ndirect}) then contracts to $n^{{\rm direct},n}_\nu\simeq
n^{{\rm inc}, n}_\nu + n^{{\rm inc}, n}_\nu(0){\cal S}_np_{{\rm surv},
  n}/(1-p_{{\rm surv}, n})$ for a nearly flat source spectrum $n^{{\rm
    inc}, n}_\nu(0)$ over the line profile. The light temperature is
then
\begin{eqnarray}
\langle T_n^{\rm core}\rangle_{\rm H}&\simeq&\frac{h}{k_{\rm B}}\frac{{\cal
    S}_n}{1-p_{{\rm surv}, n}}\left[\int
  d\nu\,\exp(-\tau_\nu)\frac{d\varphi_V(a_n,\nu)}{d\nu}\right]^{-1}\nonumber\\
&\simeq&0.0677\left(1-\frac{1}{n^2}\right)T^{1/2}\frac{{\cal
    S}_n}{1-p_{{\rm surv}, n}}\\
&\times&\left[\int
  dx\,\exp(-\tau_x)\frac{d\phi_V(a_n,x)}{dx}\right]^{-1}~{\rm
  K}.\nonumber
\label{eq:TLcore}
\end{eqnarray}
Since $d\varphi_V/d\nu<0$ blueward of the line centre, $\langle
T_n^{\rm core}\rangle_{\rm H}<0$, and the radiation field again
results in a heating term in addition to the recoil term. The term can
be large and even dominate the recoil heating term. Typical
temperatures are provided in Table~\ref{tab:supn}. The light
temperature scales like $T^{1/2}$.

The contribution to the heating from photons produced in cascades
depends entirely on integrations over the frequency redistribution
functions for the photon products following the scattering of a Ly-$n$
photon which is degraded into lower energy photons (see the
Appendix). For example, in the limit of complete redistribution,
$R_{n^\prime
  n}(\nu^\prime,\nu)=\varphi_V(a_{n^\prime},\nu^\prime)\varphi_V(a_n,\nu)$,
the contribution of direct Ly-$n^\prime$ scatters to the number
density of Ly-$n$ photons produced through cascades takes the form
$n^{{\rm direct-cascade},n}_\nu \simeq \sum_{n^\prime=n+2}^{n_{\rm
    max}}\,n^{{\rm inc}, n^\prime}_\nu(0)(\sigma_{n^\prime}/\sigma_n)
{\cal S}_{n^\prime} p_{n^\prime,n}/(1-p_{{\rm
    surv},n})$. Considerations similar to the above show that the
light temperature will be very large in magnitude for a source
spectrum $n^{{\rm inc}, n^\prime}_\nu(0)$ nearly flat across the
Ly-$n^\prime$ line profile, so that these photons will heat the gas
primarily through recoils.

The net heating rate $G_{\rm Heat}^{\rm Ly}$ by higher order Lyman
photons is shown in Figure~\ref{fig:GHeat} as a function of the
distance from a source at $z_S=8$ in an IGM with temperatures of
$T=10$~K and $T=100$~K. The steps correspond to the increasing number
of Lyman series transitions that contribute to the overall rate
through radiative cascades as the source is approached. Almost all the
heating results from direct Ly-$n$ photon scatterings; the
contribution from cascades is about three orders of magnitude
smaller. The heating rates are normalised by the local \Lya\ photon
thermalisation scattering rate $P_{\rm
  th}\simeq6.8\times10^{-12}[(1+z)/9]\,{\rm s^{-1}}$ \citep{MMR97},
required to match the scattering rate of CMB photons and so couple the
spin temperature to the kinetic temperature of the neutral hydrogen
through the W-F mechanism. (So geometric dilution is not included in
the figure:\ if $P_\alpha=P_{\rm th}$ at a particular radius, the
heating rate at smaller radii would be larger in proportion to
$1/r^2$.)  When the \Lya\ intensity is sufficiently strong to initiate
the W-F effect, the collisional heating by the scattering of higher
order Ly-$n$ photons provides a substantial increase to the
temperature of the IGM near a source. An increase in temperature
results in a reduction in the Ly-$n$ IGM optical depths (see
Figure~\ref{fig:taunu}), and a resulting increase in the heating
rate. For comparison, also shown in Figure~\ref{fig:GHeat} (dotted
line) is the radial trend of the enhanced total \Lya\ scattering rate
$P_\alpha$, including the effect of cascades from higher orders. The
additional contribution due to cascades above the incident \Lya\
photon rate is found to scale nearly like $r^{-3/7}$, although the
dependence is somewhat sensitive to the incident source spectrum. It
is very insensitive to the IGM temperature.

A fully accurate treatment of the heating rate by higher order Lyman
series photons requires solving the radiative transfer equation for
both the directly incident photons, with their rescatters, and the
photons produced through subsequent radiative cascades. It is
expected, however, that the heating rates will be reduced by at most a
few per cent., as the photons establish only partial thermal
equilibrium with the gas after only a few to several scatters
\citep{Meiksin06}, by which time they will degrade to lower energy
photons. Even if the full contribution of photons produced in
radiative cascades is excluded, the rate due to the incident higher
order Ly-$n$ photons is still substantial.

\section{The Wouthuysen-Field effect for the Lyman series}
\label{sec:WFELymann}

In this section, the Wouthuysen-Field effect is extended to the
scattering of higher order Lyman resonance line photons. Computing the
scattering rates at the hyperfine level introduces several
complicating factors over Eq.~(\ref{eq:ndirect}) and
Eq.~(\ref{eq:ncascade}). First, the redistribution functions defined
at the hyperfine level should be used. These are not straightforward,
as the ground level is no longer sharp (in particular, the decay from
the triplet to singlet states give rise to the 21cm line). Second, a
photon emitted at one hyperfine resonance may redshift into a redder
hyperfine resonance before rescattering, as $(\Delta\nu_{\rm
  hfs}/\nu_n)c\sim{\cal O}(0.1\kms)$, requiring a travel distance of
only 0.1~kpc at $z=8$. Third, for a similar reason the absorption
profiles will greatly overlap when the thermal motions of the atoms
are included. Fourth, the survival probabilities of individual
hyperfine resonance line photons must be used. Fifth, scatters in the
wings complicate the averaging over the Voigt profiles since the
natural line broadenings of individual hyperfine transitions differ
from the total line widths for a Ly-$n$ transition. This latter effect
moreover results in a light temperature in the context of the W-F
effect that differs from the light temperature of Eq.~(\ref{eq:TnH})
in the context of photon collisional heating, unlike the case for
scattering rates dominated by core scatters \citep{Meiksin06}.

To make an estimate of the magnitude of the role higher order Ly-$n$
photons play in the W-F mechanism, these effects will be neglected. In
particular, it will be assumed that the photons enter the core so that
the Voigt profile shapes for the hyperfine transitions are the same as
for their parent Ly-$n$ transitions. Since the scattering rates of
photons that scatter far in the wings are negligible, this should be a
reasonably good approximation to the overall effect. Accordingly, the
photon occupation numbers for a given hyperfine transition become the
same as for the parent Ly-$n$ transition. The hyperfine scattering
rate per atom in the lower state is then $P_n=(g_u/g_l)A_{ul}{\cal
  N}_{\nu_n}$, where $g_u$ and $g_l$ are the statistical weights of
the upper and lower hyperfine structure states, respectively, and
$A_{ul}$ is the spontaneous decay rate for the hyperfine transition.

Because the level populations above the ground state are assumed to be
excited exclusively by the scattering of Lyman resonance line photons,
in equilibrium the ratio of the occupation number of the excited
states to the ground state levels will all be of the order of the
photon occupation numbers of the incident Lyman photons or less,
depending on the branching ratios from the upper state. The $2^2S$
state is an exception. Since radiative decays from the $2^2S$ state to
the ground state are forbidden by the selection rules for dipole
transitions, this state will fill up until other processes become
effective. The dominant mechanism for vacating the state is through
two-photon decays. For this state, the occupation number will be
larger than the other excited states by a factor on the order of the
ratio of the dipole decay rate to the two-photon decay rate, still
resulting in occupation numbers much smaller than the ground state
levels because of the smallness of the photon occupation
number. Accordingly, resonance line scatterings from states above the
ground state may be neglected, as these will result in corrections to
the state populations quadratic in the photon occupation numbers of
the incident radiation field. Denoting the singlet and triplet state
occupations by $n_s$ and $n_t$, respectively, and all the hyperfine
levels above the ground state by $n_i$ with $i=1$ to $i=N$ indicating
the levels that may be reached by scattering Lyman resonance line
photons up to order $n$, the radiative cascade equations become
\begin{eqnarray}
\frac{dn_s}{dt}&=&\sum_{i=1}^N\, n_iA_{is} -
n_s\sum_{i=1}^N\,g_i{\cal N}_{si}A_{is} + P^R_{ts}n_t-P^R_{st}n_s,\nonumber\\
\frac{dn_t}{dt}&=&\sum_{i=1}^N\, n_iA_{it} -
n_t\sum_{i=1}^N\,\frac{g_i}{3}{\cal N}_{ti}A_{it}+P^R_{st}n_s-P^R_{ts}n_t,
\label{eq:cascadeground}
\end{eqnarray}
and
\begin{eqnarray}
\frac{dn_i}{dt}&=&g_iA_{is}{\cal N}_{si}n_s +
\frac{1}{3}g_iA_{it}{\cal
  N}_{ti}n_t+\sum_{j=i+1}^Nn_jA_{ji}-\nonumber\\
&&n_i\left(A_{is}+A_{it}+\sum_{j=1}^{i-1}A_{ij}\right),
\label{eq:cascade}
\end{eqnarray}
where ${\cal N}_{si}$ and ${\cal N}_{ti}$ denote the photon occupation
number between state $i$ and the singlet and triplet ground state,
respectively, $A_{ij}$ denotes the electric dipole Einstein $A$
spontaneous decay rate from hyperfine level $i$ to hyperfine level
$j$, and $g_i=2F_i+1$ denotes the degeneracy of hyperfine level $i$,
where $F_i=J_i+I$ for total angular momentum $J_i=L_i+S$, nuclear spin
$I=\pm1/2$, angular momentum $L_i=l_i$, where $l_i$ is the orbital
angular momentum of state $i$, and electron spin $S=\pm1/2$. The
transition rates are computed similarly to the transitions above, but
applying the Russell-Saunders multiplet formalism twice to allow for
$J$ and $I$ coupling to $F$. The resulting rates were checked by
verifying that applying the sum rules recovers the fine-structure and
full spontaneous decay rates.

It is convenient to label the states by increasing $n$, then by
increasing $L$ followed by increasing $J$ and $F$. Since only
spontaneous decays from levels $n$ to levels $n^\prime$ with
$n^\prime<n$ are considered, this labelling ensures the transitions
from states with higher labels to lower will always correspond to
transitions from higher energy states to lower. (It is noted that the
labelling scheme does not correspond always to increasing energy
levels. For example, the $n^{2S+1}L_J^F=2^2S_{1/2}^1$ energy level is
higher than both the hyperfine levels of $2^2P_{1/2}$,
\citet{1940ApJ....91..215B}). Since fine and hyperfine transitions
between states with the same principal quantum number are neglected,
there is no need to reorder the labelling by energy within these
quantum structure levels. If transitions up to $n=n_{\rm max}$ are
allowed, then the total number of hyperfine states above the $n=1$
states is $N=2(n_{\rm max}^2-1)$. Labelling the $2^2S_{1/2}^{0}$ state
by $i=1$ and the $2^2S_{1/2}^1$ state by $i=2$, the rates $A_{1s}$ and
$A_{2t}$ are taken to correspond to 2-photon emission at the rate
$A_{2\gamma}=8.23\,{\rm s}^{-1}$ \citep{1951ApJ...114..407S}. The
corresponding two-photon absorption terms from the ground state
singlet and triplet states to the pair of $2^2S_{1/2}$ hyperfine
states are therefore highly negligible and
excluded. Eqs.~(\ref{eq:cascadeground}) also include radiative
excitation and de-excitation between the singlet and triplet hyperfine
states, given by the respective rates $P^R_{st}$ and $P^R_{ts}$, to
allow for an incident continuum radiation field due to the CMB or
nearby radio sources. Similar collisional terms may be added if
required.

The excited levels $n>1$ will rapidly achieve statistical equilibrium
on a timescale of order $A_{ij}^{-1}\sim{\cal O}(10^{-8}\,{\rm s})$
(except for the $2^2S_{1/2}$ states, for which the timescale is on the
order of 0.1~s). The states may thus be assumed to have reached a
steady state, so that Eq.~(\ref{eq:cascade}) reduces to a matrix
equation. It is convenient to renormalise the level occupations by
$\tilde n_i=n_i/[n_s{\cal N}_L^{\rm inc}(0)]$, where ${\cal N}_L^{\rm
  inc}(0)$ is the incident photon occupation number just longward of
the Lyman edge, and to define the ratio $r=n_t/n_s$ between the
triplet and singlet ground state levels. Denoting the rescaled
occupation levels by the vector ${\bf \tilde n}$, the matrix equation
for the excited levels becomes
\begin{equation}
{\bf M}\cdot{\bf \tilde n}={\bf y},
\label{eq:matrix}
\end{equation}
where {\bf M} is an upper triangular matrix with diagonal elements
$M_{ii}=A_{is}+A_{it}+\sum_{j=1}^{i-1}A_{ij}$, $M_{ij}=-A_{ji}$ and
$\bf y$ is an absorption vector with elements $y_1=y_2=0$ (indicating
no 2-photon absorption to states $i=1$ and $i=2$), and
$y_i=g_iA_{is}v_{si} + (g_i/3)A_{it}rv_{ti}$ for $i>2$. Here the
notation $v_{si}={\cal N}_{si}/{\cal N}_\alpha^{\rm inc}(0)$ and
$v_{ti}={\cal N}_{ti}/{\cal N}_\alpha^{\rm inc}(0)$ has been
introduced.

Following \citet{1958PROCIRE.46..240F}, the value for $r$ may be
determined by casting Eq.~(\ref{eq:cascadeground}) in the form
\begin{eqnarray}
\frac{dn_s}{dt}&=&P^{\rm Ly}_{ts}n_t-P^{\rm Ly}_{st}n_s
+P^R_{ts}n_t-P^R_{st}n_s,\nonumber\\
\frac{dn_t}{dt}&=&P^{\rm Ly}_{st}n_s-P^{\rm Ly}_{ts}n_t
+P^R_{st}n_s-P^R_{ts}n_t,
\label{eq:ground}
\end{eqnarray}
where $P^{\rm Ly}_{st}$ and $P^{\rm Ly}_{ts}$ are the effective
excitation and de-excitation rates of the triplet state by the
scattering of Lyman resonance line photons. Defining $T_{st}$ as the
effective temperature corresponding to the 21-cm transition, given by
$hc/(k_B\lambda_{\rm 21cm})=T_{st}\simeq0.068$~K, and the 21-cm
spontaneous transition rate $A_{ts}=2.85\times10^{-15}\,{\rm s}^{-1}$
\citep{1952ApJ...115..206W}, in equilibrium, $r=3\exp(-T_{st}/T_S)$,
where $T_S$ is the spin temperature, is given by $r=(P^{\rm
  Ly}_{st}+P^R_{st})/ (P^{\rm Ly}_{ts}+P^R_{ts})$, where
$P^R_{st}=3A_{ts}(T_R/T_{st})$ and
$P^R_{ts}=A_{ts}(1+T_R/T_{st})$. Here $T_R$ is the brightness
temperature of the incident continuum radiation field at the 21cm
transition frequency. In statistical equilibrium without an incident
continuum ($T_R=0$), $r=3\exp(-T_{st}/T_L)=P_{st}^{\rm Ly}/P_{ts}^{\rm
  Ly}$, where $T_L$ is defined as the light temperature, and is
identical (to order $k_{\rm B}T_{st}/h\nu_\alpha$) to the harmonic
mean temperature Eq.~(\ref{eq:TnH}) of the radiation field when the
scattering rate is dominated by core scatters \citep{Meiksin06}. It
relaxes to the kinetic temperature of the gas after a few dozen to a
few hundred scatters across the resonance line centre
\citep{1959ApJ...129..551F, Meiksin06}. Using the definition of $T_L$
above, the spin temperature may be expressed generally as
\begin{equation}
  T_S=T_{st}\Biggl[\frac{T_{st}}{T_L}+\log\frac{T_{st}
\left(1+T_R/T_{st}\right)/T_L+y_{\rm Ly}}
{T_R\exp(T_{st}/T_L)/T_L+y_{\rm Ly}}\Biggr]^{-1},
\label{eq:TS}
\end{equation}
where $y_{\rm Ly}=T_{st}P_{ts}^{\rm Ly}/(T_LA_{ts})$. (The expression
must be modified if electron or atomic collisions are important, here
assumed negligible.)

When the light temperature of the \Lya\ photons has relaxed to the
kinetic temperature of the gas, the problem is entirely determined
once the triplet de-excitation rate $P_{ts}^{\rm Ly}$ is
specified. For pure \Lya\ scattering, $P_{ts}^{\rm Ly} =
(4/27)P_\alpha$, where $P_\alpha=3{\cal N}_\alpha A_\alpha$
\citep{1958PROCIRE.46..240F}. The higher order generalisation is given
by
\begin{eqnarray}
  P_{ts}^{\rm Ly}&=&\sum_{nlJF}\left(\frac{2F+1}{3}\right){\cal
    N}_{t, (n,l,J,F)}\nonumber\\
  &\times& A(n,l,J,F;1,0,\frac{1}{2},1) p_s(n,l,J,F),
\label{eq:PTSit}
\end{eqnarray}
where ${\cal N}_{t, (n,l,J,F)}$ is the occupation number of photons
with frequencies at the hyperfine structure resonance between the
ground state triplet state and the state $(n,l,J,F)$, and
$p_s(n,l,J,F)$ is the probability that an electron in the hyperfine
state $(n,l,J,F)$ produces a decay to the ground singlet state,
\begin{eqnarray}
  p_s(n,l,J,F)&=&A_{nlJF}^{-1}\sum_{n^\prime<n,l^\prime,J^\prime,F^\prime}
A(n,l,J,F;n^\prime,l^\prime,J^\prime,F^\prime)\nonumber\\
&\times& p_s(n^\prime,l^\prime,J^\prime,F^\prime).
\label{eq:PShfs}
\end{eqnarray}
Here, $A_{nlJF}=\sum_{n^\prime<n,l^\prime,J^\prime,F^\prime}
A(n,l,J,F;n^\prime,l^\prime,J^\prime,F^\prime)$ ($=A_{n,l}$ by the sum
rules), is the total decay rate of state $(n,l,J,F)$. The
probabilities $p_s$ may be solved for iteratively from
Eq.~(\ref{eq:PShfs}), initiated by $p_s(2,0,\frac{1}{2},0)=1$,
$p_s(2,0,\frac{1}{2},1)=0$, $p_s(2,1,\frac{1}{2},0)=0$,
$p_s(2,1,\frac{1}{2},1)=1/3$, $p_s(2,1,\frac{3}{2},1)=2/3$, and
$p_s(2,1,\frac{3}{2},2)=0$.

The solution to the equations shows the singlet-triplet excitation and
triplet-singlet de-excitation rates are boosted only by the increase
in the \Lya\ collision rate $P_\alpha$ due to the \Lya\ photons
produced in cascades following the scattering of higher order Ly-$n$
photons (Figure~\ref{fig:GHeat}). The ratio $P_{ts}^{\rm
  Ly}=(4/27)P_\alpha$ is found to be accurate to a small fraction of
one per cent. even close to a source (down to a proper separation
smaller than 20~kpc for a source at $z_S=8$).

\section{Astrophysical consequences}
\label{sec:consequences}

The redshift $z_r$ of the reionisation of the IGM is constrained
primarily by measurements of intergalactic \Lya\ absorption in the
spectra of high redshift QSOs and by polarisation measurements of the
CMB. The spectra of high redshift QSOs constrain $z_r>5.7$, above
which the \Lya\ optical becomes immeasurably large
\citep{2006ARA&A..44..415F}. The five-year {\it Wilkinson Microwave
  Anisotropy Probe} polarisation data yield $z_r=11.0\pm1.4$ assuming
sudden reionisation, with $2\sigma$ and $3\sigma$ lower limits of
$z_r>8.2$ and $z_r>6.7$, respectively \citep{2009ApJS..180..306D}.
Radio telescope efforts to detect the EoR through its 21cm signature
are focussed on appropriate redshift ranges accordingly:\ $6.1<z<10.8$
(LOFAR), $z<17$ (MWA), $6.1<z<27$ (PaST/21CMA) and $6.1<z<13$
(PAPER). Since the impact of higher order Lyman photons on the IGM is
sensitive to redshift, the astrophysical consequences are illustrated
for a range of redshifts.

\subsection{Photon collisional heating}

\begin{figure}
\includegraphics[width=3.3in]{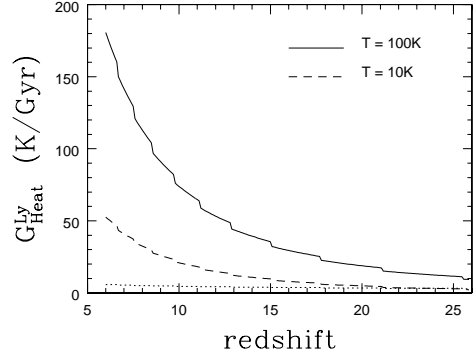}
\caption{Lyman photon collisional heating rate at a distance of
  100~kpc from a $10\,\msun\,{\rm yr^{-1}}$ starburst, for source
  redshifts in the range $6<z_S<26$. Shown for IGM temperatures of
  $T=100$~K (solid line) and $T=10$~K (dashed line). Also shown is the
  corresponding total \Lya\ collision rate $P_\alpha/P_\alpha^{\rm
    inc}(0)$, normalised by the collision rate of the incident
  photons.
}
\label{fig:GHeatz}
\end{figure}

The impact Lyman photon collisional heating may have on the
temperature of the IGM is illustrated in Figure~\ref{fig:GHeatz} using
a starburst galaxy. A galaxy continuously forming low metallicity
($Z=0.05Z_\odot$) stars at the rate $10\,\msun\,{\rm yr}^{-1}$ with
masses between $1<M<100\msun$ and a Salpeter IMF will have a steady
luminosity at $\lambda=915\,{\rm \AA}$ of
$L_\nu=3.8\times10^{28}\,{\rm erg\,s^{-1}\,Hz^{-1}}$ after about
$10^7$~yrs \citep{1999ApJS..123....3L}. This corresponds to a photon
occupation number at the Lyman edge of ${\cal
  N}_\nu\simeq4.8\times10^{-23} r_{\rm Mpc}^{-2}$, where the (proper)
distance is normalised to 1~Mpc.

The heating rate is found to decline with increasing source redshift
as the IGM becomes increasingly optically thick to Lyman photon
scattering, reducing the effective Lyman photon collision rates. A
second reduction factor with redshift is the decrease in the number of
Lyman orders that may contribute at a fixed proper distance before a
photon redshifts into the resonance frequency of the next lower
order. Each step in Figure~\ref{fig:GHeatz} corresponds to a decrease
by one order. For comparison, also shown in Figure~\ref{fig:GHeatz}
(dotted line) is the redshift trend of the enhanced total \Lya\
scattering rate $P_\alpha/P_\alpha^{\rm inc}(0)$, including the
contribution from cascades following the scattering of higher order
Lyman photons. The rate is normalised by the incident rate excluding
the cascade contribution.

Because the optical depth of the IGM diminishes with increasing
temperature, the photon collisional heating rate strengthens with
increasing temperature, varying as $G_{\rm Heat}^{\rm Ly}\propto
T^{1/2}$. This gives rise to a weak thermal instability that grows
quadratically with time. Photon collisional heating may thus produce a
substantial increase in the temperature of the IGM surrounding a
source and transform an absorption signature against the background
CMB into an emission signature \citep{MMR97}.

\subsection{W-F effect in the absence of \Lya\ photons}

It was found that the radiative cascades following the scattering of
Ly-$n$ photons will enhance the number density of \Lya\ photons by as
much as an order of magnitude near a source. Near a galactic source,
however, dust may be present. It is unclear that \Lya\ photons will be
able to survive sufficiently long to build up the photon density
required to initiate the W-F effect before being absorbed by dust
grains. Observational searches for dust surrounding galaxies suggest
dust may be present at least as far out as 100~kpc around low redshift
galaxies \citep{1994AJ....108.1619Z, 2009arXiv0902.4240M}. If
comparable amounts of dust were introduced into the surroundings of
galaxies at the time of the EoR following bursts of star formation,
then the long total path \Lya\ photons must travel to migrate from the
blue frequencies at which they become trapped in the gas at large
distances into the Doppler core, where they become effective
scatterers, may exceed the mean free path for dust scattering. The
measurement of \citet{ 2009arXiv0902.4240M} of
$A_V\simeq2.4\times10^{-3}(r_p/100h^{-1}\,{\rm kpc})^{-0.84}$ at a
projected separation of $r_p$ suggests a mean free path for colliding
with dust grains of about $4\times10^5(r/100h^{-1}\,{\rm
  kpc})^{1.74}$~kpc a distance $r$ from the centre of the galaxy,
assuming $A_\lambda\simeq1.086N_dQ_e\sigma_d$ for a column density of
$N_d$ dust grains with cross sections $\sigma_d$ and an extinction
efficiency factor of $Q_e\simeq2$ in the visual
\citep{1978ppim.book.....S}. For \Lya\ photons with a mean free path
on the order of the distance to which they are able to survive before
becoming trapped within the blue wing of an atom, the distance
travelled by the photons (of order $N_{\rm scat}\simeq1000$ in the
wings times the \Lya\ mean free path), before diffusing into the
resonance line core approaches the mean free path for scattering off a
dust grain, which may cause some suppression of the W-F effect. The
density of dust around starbursts, however, is likely to be
considerably larger than in a normal galaxy, by as much as two orders
of magnitude, and may extend to quite large radii
\citep{2000ApJS..129..493H}. In this case the diffuse \Lya\ photon
radiation field required to drive the W-F effect may become
substantially suppressed. Only the photons that have scattered too few
times to have migrated far into the core would contribute.

\begin{figure}
\includegraphics[width=3.3in]{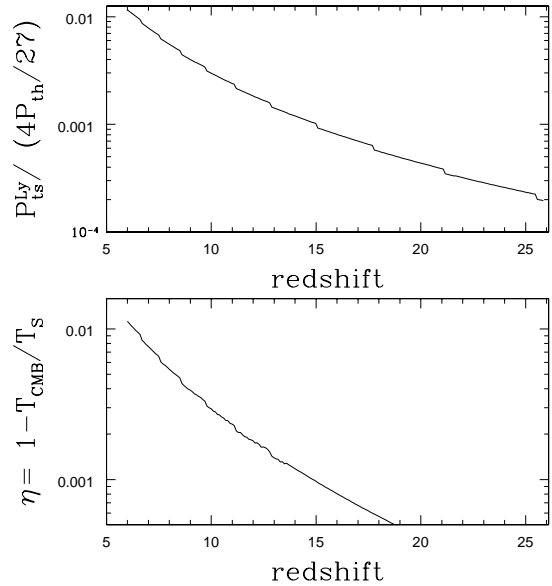}
\caption{(Upper panel)\ The de-excitation rate of the ground triplet
  state, normalised by the rate corresponding to the thermalisation
  scattering rate $P_{\rm th}$, induced by higher order Lyman series
  photons in the absence of \Lya\ photons. The rate is shown at a
  distance of 100~kpc from a $100\,\msun\,{\rm yr^{-1}}$ starburst,
  for source redshifts in the range $6<z_S<26$. An IGM temperature
  $T=10$~K is assumed. (Lower panel)\ The corresponding
  21cm radiation efficiency $\eta$.
}
\label{fig:PSTTS_etaz}
\end{figure}

\begin{figure}
\includegraphics[width=3.3in]{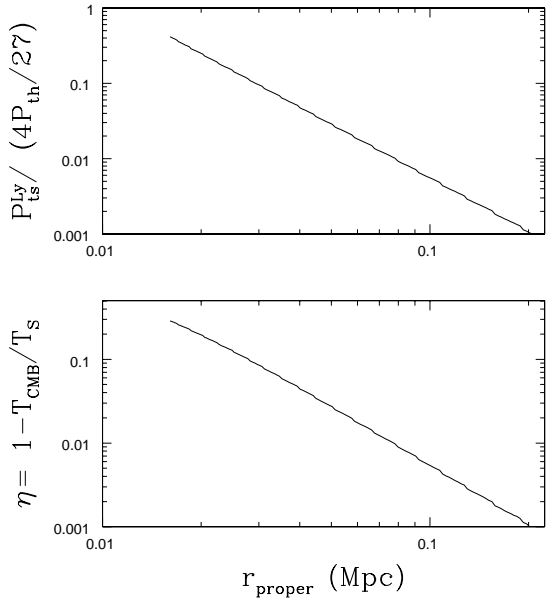}
\caption{(Upper panel)\ The de-excitation rate of the ground triplet
  state, normalised by the rate corresponding to the thermalisation
  scattering rate $P_{\rm th}$, induced by higher order Lyman series
  photons in the absence of \Lya\ photons. The rate is shown for a
  $100\,\msun\,{\rm yr^{-1}}$ starburst at $z_S=8$. An IGM temperature
  $T=10$~K is assumed. (Lower panel)\ The corresponding 21cm radiation
  efficiency $\eta$.
}
\label{fig:PSTTS_etar}
\end{figure}

The loss of the \Lya\ photons, however, does not mean that the W-F
effect will be extinguished. Although far fewer in number, the higher
order Lyman series photons will still induce a W-F effect. Because the
higher order photons will not achieve thermal equilibrium with the
gas, the spin temperature will no longer reflect the kinetic
temperature of the gas but rather the net exchange rate between the
ground state singlet and triplet levels resulting from collisions by
the higher order Lyman photons.

The evolution of the triplet-singlet de-excitation rate at a (proper)
distance of 100~kpc from a $100\,\msun\,{\rm yr}^{-1}$ starburst (as
above), is shown in Figure~\ref{fig:PSTTS_etaz} (upper panel) for an
IGM temperature of $T=10$~K. (The results are found to be very
insensitive to temperature.) The de-excitation rate is only a fraction
of the rate for a total Lyman photon scattering rate matching the
thermalisation rate $P_{\rm th}$, but is non-negligible. The light
temperature $T_L$ is found to be high, varying only moderately with
redshift, from $T_L=170$~K at $z=6$ to $T_L=200$~K at $z=26$. The spin
temperature established, however, must include the scattering of CMB
photons. Shown in the lower panel is the resulting 21cm radiation
efficiency factor $\eta=1-T_{\rm CMB}/T_S$, which controls the
strength of the 21cm signal through $\delta T\simeq\eta
T_S(1-e^{-\tau})/(1+z)$, where $\delta T$ is the change in the antenna
temperature compared with the CMB temperature in a medium with 21cm
optical depth $\tau$ \citep{MMR97}. Typical efficiency factors of only
one per cent. or smaller result, so that a detection would be
difficult with the upcoming telescopes. The efficiency, however, rises
at smaller radii, as shown in Figure~\ref{fig:PSTTS_etar} for the
galaxy at $z_S=8$. It is noted the timescale for the Lyman resonance
photons to affect the spin temperature will be on the order of
$(P_{ts}^{\rm Ly})^{-1}\simeq1$~Myr, so that bursts of shorter
duration may not have much effect. A brighter source, however, would
increase the signal proportionally. Also, scatters by the residual
\Lya\ photons still outside the resonance line core, not included in
the discussion here, will contribute as well, further enhancing the
signal.

Although the emphasis in this paper is on the effect of higher order
Lyman photons around a single source, a diffuse radiation field will
result in regions sufficiently near sources that higher order photons
from multiple sources are able to penetrate them before scattering
through a lower order Lyman resonance. In this case the collective
incident radiation field from the sources is given by
\begin{equation}
  J_\nu=\frac{1}{4\pi}\int_z^\infty\,dz^\prime\,\frac{dl_p}{dz^\prime}
\frac{(1+z)^3}{(1+z^\prime)^3}\langle\epsilon_{\nu^\prime}(z^\prime)
\exp[-\tau_\nu(z,z^\prime)]\rangle,
\label{eq:Jnu}
\end{equation}
\citep{2007arXiv0711.3358M}, for a source emissivity
$\epsilon_{\nu^\prime}$, where $\nu^\prime=\nu(1+z^\prime)/(1+z)$, and
$\tau_\nu(z,z^\prime)$ is the total optical depth due to all the Lyman
resonances (and any other contributing terms) encountered by the
photons from gas between the redshifts $z$ and $z^\prime$. The
indicated spatial average is over the sources and the IGM optical
depth jointly, as these are likely correlated since sources form in
overdense regions.

\subsection{Cascade radiation emission lines}

A consequence of the scattering of Lyman series photons is the
generation of a host of intermediate emission lines produced in the
cascades following the scattering. The emissivities for these
transitions may be computed from the state occupancies given by the
solution to Eq.~(\ref{eq:matrix}).

\begin{table}
  \caption{Scaled emission line photon number emissivities ${\tilde
      \epsilon}_{nn^\prime}= {\tilde n}_nA_{nn^\prime}$ (in units of
    ${\rm s^{-1}}$), for a source at redshift
    $z_S=8$ and a $T=10$~K IGM, at (proper)
    distances from the source of 20~kpc (cols 2--3), 100~kpc (cols 4--5)
    and 1~Mpc (cols 6--7). The results are shown in pairs of Balmer and
    Paschen lines with upper principal quantum number $n$.
  }
\begin{center}
\begin{tabular}{|c|l|l|l|l|l|l|} \hline
\hline 
$n$ & ${\tilde{\epsilon}}_{n2}$ & ${\tilde{\epsilon}}_{n3}$
& ${\tilde{\epsilon}}_{n2}$ & ${\tilde{\epsilon}}_{n3}$ 
& ${\tilde{\epsilon}}_{n2}$ & ${\tilde{\epsilon}}_{n3}$ \\
\hline
3    &360953. & $\dots$ & 203484. & $\dots$ & 91735.4 & $\dots$ \\
4    &112580. & 56787.4 & 70098.6 & 33785.3 & 39451.5 & 16654.2 \\
5    &58433.7 & 27836.5 & 39844.4 & 17841.3 & 26111.4 & 10273.5 \\
6    &41193.9 & 18287.5 & 30835.2 & 12820.4 & 23025.4 & 8617.90 \\
7    &32874.8 & 13850.3 & 26252.1 & 10412.6 & 21148.9 & 7724.69 \\
8    &28240.0 & 11450.3 & 23617.6 & 9085.13 & 19960.5 & 7197.96 \\
9    &25390.9 & 10009.5 & 21964.1 & 8278.14 &  & \\
10   &23509.6 & 9077.03 & 20854.9 & 7750.76 &  & \\
11   &22200.0 & 8439.03 & 20072.8 & 7387.03 &  & \\
12   &21252.7 & 7984.15 & 19501.7 & 7126.15 &  & \\
13   &20542.8 & 7647.83 & 19068.8 & 6931.77 &  & \\ 
14   &19996.7 & 7392.26 & 18731.9 & 6782.75 &  & \\  
15   &19568.0 & 7193.75 & 18464.0 & 6665.82 &  & \\  
16   &19225.1 & 7036.52 & 18246.1 & 6571.86 &  & \\  
17   &18946.5 & 6909.93 & 18065.2 & 6494.71 &  & \\  
18   &18717.1 & 6806.52 & 17912.9 & 6430.56 &  & \\ 
19   &18525.8 & 6720.97 &  &&&\\
20   &18364.2 & 6649.25 &  &&&\\
21   &18227.6 & 6588.95 &  &&&\\
22   &18109.9 & 6537.39 &  &&&\\
23   &18008.3 & 6493.12 &  &&&\\
24   &17919.5 & 6454.62 &  &&&\\
25   &17841.7 & 6421.08 &  &&&\\
26   &17772.5 & 6391.43 &  &&&\\
27   &17710.7 & 6365.05 &  &&&\\
28   &17655.4 & 6341.53 &  &&&\\
29   &17604.7 & 6320.09 &  &&&\\
30   &17558.5 & 6300.60 &  &&&\\
31   &17515.8 & 6282.66 &  &&&\\
\hline 
\end{tabular}
\end{center}
\label{tab:emiss}
\end{table}

The scaled Balmer and Paschen photon number emissivities,
${\tilde\epsilon}_{n2}={\tilde n}_nA_{n2}$ and
${\tilde\epsilon}_{n3}={\tilde n}_nA_{n3}$, summed over the hyperfine
structure lines for a given upper principal quantum number $n$, are
shown in Table~\ref{tab:emiss} for a source at $z_S=8$ at (proper)
distances of 20~kpc, 100~kpc and 1~Mpc from the source, for an IGM at
temperature $T=10$~K. (The emissivities are very insensitive to
temperature. The values change by no more than a few per cent. for
$T=100$~K.) The bolometric emissivities in physical units are then
given by $\epsilon_{ij}=(n_{\rm H}/4)h\nu_{ij}{\cal N}_L^{\rm
  inc}(0){\tilde\epsilon}_{ij}$, where the factor of 4 takes into
account that only $1/4$ of the ground state hydrogen atoms are in the
hyperfine singlet state (see \S~\ref{sec:WFELymann}). The emissivities
${\tilde\epsilon}_{ij}$ for a source at $z_S=20$ are about a factor 5
smaller. The emissivities are found to diminish slowly with distance
from the source. The lower order Balmer sequence emissivities scale to
better than 8 per cent. accuracy as $r^{-1/3}$ for H$\alpha$,
$r^{-1/4}$ for H$\beta$ and $r^{-1/5}$ for H$\gamma$. These scalings
apply to the Paschen series Pa$\alpha$, Pa$\beta$ and Pa$\gamma$ as
well. The halo of any given emission line vanishes beyone the horizon
of the lowest order Ly-$n$ required to generate it.

The bolometric intensity a projected distance $b$ from a source at
redshift $z_S$ for a bolometric emissivity varying as
$\epsilon_{ij}=\epsilon_{ij}^*(r/r_*)^{-\alpha-2}$ is
\begin{eqnarray}
  i_{ij}&\simeq&\frac{1}{2\pi}\epsilon_{ij}^*
  r_*^{\alpha+2}(1+z_S)^{-4}
  \int_0^\infty dl\,  (b^2+l^2)^{-(\alpha+2)/2}\\
  &=&\frac{1}{2\pi^{1/2}\alpha}\frac{\Gamma(\frac{1+\alpha}{2})}
  {\Gamma(\alpha/2)}\epsilon_{ij}^*
  r_*\left(\frac{r_*}{b}\right)^{1+\alpha}
  (1+z_S)^{-4},
\label{eq:itheta}
\end{eqnarray}
where the redshift factor has been introduced to convert the bolometric
intensity to the value measured at $z=0$. The corresponding specific
intensity is
\begin{equation}
i_\nu =
\frac{1}{4\pi}\frac{c}{H(z)}\frac{\epsilon_{ij}^*}{\nu_{ij}}\frac{1}{(1+z_S)^3}
\Biggl[\left(\frac{b}{r_*}\right)^2+\left(\frac{\Delta
    v}{H(z)r_*}\right)^2\Biggr]^{-(\alpha+2)/2},
\label{eq:inu}
\end{equation}
where $\Delta v$ is the velocity difference from line centre. A
homogeneous and isotropic medium around the source has been
assumed. More realistic line profiles would need to include the
structure of the underlying density and velocity fields.

\begin{figure}
\includegraphics[width=3.3in]{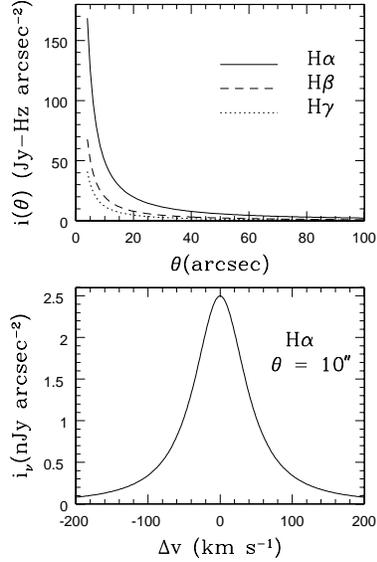}
\caption{Cascade radiation emission line profiles around a
  $100\msun\,{\rm yr}^{-1}$ starburst galaxy at $z_S=8$ in an IGM with
  temperature $T=10$~K. (Upper panel)\ Shown are bolometric H$\alpha$,
  H$\beta$ and H$\gamma$ intensities, corresponding to the respective
  observed wavelengths of 5.9$\mu$m, 4.4$\mu$m and 3.9$\mu$m. The
  integrated fluxes within one arcminute of the source are $f_{{\rm
      H}\alpha}=1.6\times10^{-18}\,{\rm erg\,cm^{-2}\,s^{-1}}$,
  $f_{{\rm H}\beta}=8.2\times10^{-19}\,{\rm erg\,cm^{-2}\,s^{-1}}$ and
  $f_{{\rm H}\gamma}=5.8\times10^{-19}\,{\rm erg\,cm^{-2}\,s^{-1}}$.
  (Lower panel)\ H$\alpha$ specific intensity at $\theta=10$~arcsec.
}
\label{fig:iprofile}
\end{figure}

The profiles may be illustrated by the cascade line radiation produced
by a $100\msun\,{\rm yr}^{-1}$ starburst at $z_S=8$ in a $T=10$~K
IGM. The bolometric intensities for H$\alpha$, H$\beta$ and H$\gamma$
are then $i_{{\rm H}\alpha}\simeq1.1\times10^{-20}\theta^{-4/3}\, {\rm
  erg\,cm^{-2}\,s^{-1}\,arcsec^{-2}}$, $i_{{\rm
    H}\beta}\simeq4.3\times10^{-21} \theta^{-5/4}\,{\rm
  erg\,cm^{-2}\,s^{-1}\,arcsec^{-2}}$ and $i_{{\rm
    H}\gamma}\simeq2.6\times10^{-21}\theta^{-6/5}\, {\rm
  erg\,cm^{-2}\,s^{-1}\,arcsec^{-2}}$, at an observed angular
separation from the source $\theta$ measured in arcsecs. The profiles
are shown in the upper panel of Figure~\ref{fig:iprofile}. In the
lower panel is shown the specific intensity at $\theta=10$~arcsec,
$i_\nu\simeq2.5\times10^{-32}\,{\rm
  erg\,cm^{-2}\,s^{-1}\,Hz^{-1}\,arcsec^{-2}} [1+(\Delta v/47\,{\rm
  km\,s^{-1}})^2]^{-7/6}$.

Although the intensities are small, the integrated fluxes from the
emission line halos around the source are appreciable. At $z_S=8$,
Ly$\beta$ photons are able to travel as far 56~Mpc from the source
before redshifting into the \Lya\ resonance. This corresponds to an
angular distance of 3.4$^\circ$. The total integrated flux is $f_{{\rm
    H}\alpha}^{\rm tot}\simeq5.5\times10^{-17}\,{\rm
  erg\,cm^{-2}\,s^{-1}}$. Whilst isolating the flux from such an
extended halo is likely unfeasible, even within the inner arcminute
the total flux is $f_{{\rm
    H}\alpha}(1^\prime)\simeq1.6\times10^{-18}\,{\rm
  erg\,cm^{-2}\,s^{-1}}$. Similarly, the fluxes for H$\beta$ and
H$\gamma$ are $f_{{\rm H}\beta}(1^\prime)\simeq8.2\times10^{-19}\,{\rm
  erg\,cm^{-2}\,s^{-1}}$ and $f_{{\rm
    H}\gamma}(1^\prime)\simeq5.8\times10^{-19}\,{\rm
  erg\,cm^{-2}\,s^{-1}}$. An observing strategy involving narrow band
near infra-red imaging over a wide area would be required to detect
the halos. The existence of multiple emission lines has the advantage
that it would be possible to correlate the signals in frequency space,
with continuum emission from contaminating sources subtracted. For
example, defining $\delta f_{{\rm H}\beta}=f_{{\rm H}\beta}+f_{\rm
  cont}-\langle f_{\rm cont}\rangle$, and similarly for $H\gamma$, the
correlation of the H$\beta$ and H$\gamma$ signals is $\langle \delta
f_{{\rm H}\beta} \delta f_{{\rm H}\gamma}\rangle = \langle f_{{\rm
    H}\beta}f_{{\rm H}\gamma}\rangle + [\langle f_{\rm
  cont}^2\rangle-\langle f_{\rm cont}\rangle^2]$, so that the signals
only need to be detectable above the noise of the continuum across the
frequency band of the detector. The emission lines will also give rise
to a diffuse sky brightness. At a few such bright sources per square
degree, a minimum near infra-red bolometric sky brightness of
$\sim0.1$~mJy~sr$^{-1}$ may be expected; the value would be much
larger if a much higher density of sources was required to reionise
the IGM. Although the detection of individual hyperfine emission lines
is unlikely in the near future, it is noted that the state occupancies
at the hyperfine structure level are found to scale nearly in
proportion to their statistical weights $2F+1$.

It was assumed in \S~\ref{sec:WFELymann} that the optical depths of
all levels above the ground $n=1$ levels were negligible so that
absorptions from them could be ignored. This assumption is now
justified. As discussed in \S~\ref{sec:WFELymann}, the ratio of the
occupancies to the ground state occupancy will be on the order of the
photon occupation numbers, and typically much smaller. Only the $2
^2S_{1/2}$ states have large values, of $10^3-10^5$, because of the
smallness of the two-photon decay rate.

The optical depth for transitions between hyperfine states $i$ and $j$
is (for absorption from $i$ to $j$),
\begin{eqnarray}
  \tau_{ij}&=&\lambda_{ij}\sigma_{ij}n_i(z)H(z)^{-1}\\
  &\simeq&1549\frac{g_j}{g_i}\left(\frac{\lambda_{ij}^3A_{ji}}{\lambda_\alpha^3
      A_\alpha}\right){\tilde n}_i{\cal N}_L^{\rm
    inc}(0)(1+z)^{3/2}\nonumber
\label{eq:tauhfs}
\end{eqnarray}
where $\lambda_{ij}^3A_{ji}$ has been normalised by the value for
\Lya. Even for the $2 ^2S_{1/2}$ hyperfine states ($n_1$ and $n_2$ in
the notation of this paper), the optical depths will be
tiny. Balancing $n_1 A_{2\gamma}\sim{\cal N}_\alpha^{\rm
  inc}(0)A_\alpha n_s$ gives an upper limit ${\tilde n}_1\sim{\cal
  O}(A_\alpha/A_{2\gamma}\simeq8\times10^7)$; in practice the value is
more than two orders of magnitude smaller because of the suppression
factors ${\cal S}_n$ and reduced cross-sections for $n>2$. Since
generally ${\tilde n}_i<{\tilde n}_1$, using this generous upper limit
for ${\tilde n}_1$ shows that $\tau_{ij}<1$ provided
\begin{equation}
\frac{g_j}{g_i}\left(\frac{\lambda_{ij}^3A_{ij}}{\lambda_\alpha^3
    A_\alpha}\right) < 8.48\times10^{-12}\left[{\cal N}_\alpha^{\rm
    inc}(0)\right]^{-1}(1+z)^{-3/2},
\label{eq:Aijlimit}
\end{equation}
an equality that is readily met for any relevant transition and
realistic radiation field. As an illustration, for a source at $z_S=8$
in a $T=10$~K IGM, at a distance of 100~kpc from the source the
largest $n\ne n^\prime$ hyperfine transition optical depth is found
for absorption from
$(n,l,J,F)=(2,0,\frac{1}{2},1)\rightarrow(3,1,\frac{3}{2},2)$ with
$\tau\simeq1.3\times10^{10}{\cal N}_\alpha^{\rm
  inc}(0)\simeq5\times10^{-11}$ for ${\cal N}_\alpha^{\rm
  inc}(0)\simeq3.5\times10^{-21}$ corresponding to the thermalisation
\Lya\ scattering rate $P_{\rm th}$ at $z=8$.

\section{Conclusions}

Higher order Lyman series photons from sources at high redshift may
have a substantial influence on the 21cm signature from the IGM in the
vicinity of the source before the IGM is reionised. Taking into
account the diminishing cross section for scattering with increasing
order, it is found that Ly-$n$ resonance line photons with $n\ge5$ or
6 will first scatter in the resonance line core, where they scatter
efficiently. The consequences for the subsequent rate of collisional
heating, the intensity of the Wouthuysen-Field effect, and the
formation of extended diffuse halos of emission lines produced in
radiative cascades are summarised.

Whilst \Lya\ photons readily establish thermal equilibrium with the
IGM, higher order resonance line photons degrade into lower order
photons before they are able to do so. Because of the much smaller
fraction of higher order Ly-$n$ photons scattered out of a line of
sight compared with \Lya, they are able to produce a substantial
amount of collisional heating. The heating arises primarily from the
blue distortion of the photon density distribution, as only photons
blueward of the resonance line frequency are able to survive to large
distances without being scattered out of the line of sight. Typical
light temperatures for Ly-$n$ photons, where $n$ is the principal
quantum number of the upper state, of $\langle T_n\rangle_{\rm H}\approx
-0.015T^{1/2}$~K are obtained, where $T$ is the temperature of the
IGM. Resulting collisional heating rates of several tens to hundreds
of Kelvin per Gyr are obtained. The heating rate scales as $T^{1/2}$,
inducing a weak thermal instability in the IGM, with the IGM
temperature increasing quadratically with time. One consequence would
be that warm regions are heated more rapidly than cooler regions,
which would contribute to the patchiness of a 21cm absorption
signature against the CMB where the spin temperature is still less
than the temperature of the CMB.

Radiative cascades following the scattering of higher order Lyman
photons will add substantially to the number density of \Lya\ photons
near a source, enhancing their number by an order of magnitude. The
number of \Lya\ photons produced in cascades is found to be as much as
30 per cent. higher than previous estimates.

Measurements of the dust content in nearby starburst galaxies suggest
the W-F effect induced by \Lya\ photons may be partially suppressed
due to the absorption of \Lya\ photons by dust grains, and possibly
even completely eliminated, in the vicinity of starburst galaxies at
high redshift during the EoR if comparable amounts of dust were
present. It is demonstrated that higher order Lyman series photons
will induce a non-negligible W-F effect even in the complete absence
of \Lya\ photons. The 21cm radiation efficiency at distances exceeding
100~kpc from the galaxy would then be one per cent. or smaller except
for extremely strong bursts, although the signal is stronger closer to
the galaxy. The signal would still likely be too weak to be discovered
around an individual galaxy by first generation 21cm detection
experiments, although it may be detectable by a Square Kilometre
Array. Future radio detections and upper limits may provide a means of
exploiting the suppression to constrain the dust production rates and
transport efficiency in high redshift galaxies.

Cascade radiation following the scattering of higher order Lyman
photons will produce emission line halos extending over several
arcminutes around a source, limited only by the distance to which
Lyman photons may travel before redshifting into the resonance
frequency of the next lower order. It is shown that the integrated
lower order Balmer fluxes within one arcminute surrounding starburst
galaxies would be substantial. The detection of the halos would
provide a unique means of confirming that candidate reionisation
sources are in fact surrounded by an IGM that is still largely
neutral.

The effects discussed here are in the context of a homogeneous and
expanding universe. Accounting for cosmological structures will modify
the results in several ways \citep{2009MNRAS.393..949H}. The local
density and velocity fields will affect the mean free paths of the
Lyman photons, as will heat input from other possible sources such as
Active Galactic Nuclei or Quasi-Stellar Objects. Of particular
interest are \HII\ regions surrounding photoionisation sources. The
horizons of high order Lyman photons could then be substantially
extended as the ionised gas would become optically thin at their
resonance frequencies. The collisional heating, W-F effect, and the
production of emission line halos would then extend into the neutral
gas just beyond the ionisation front. Even with these complications,
it is expected that the salient features of the effects discussed here
will still be present.

\begin{appendix}
\section{Redistribution function for degraded Lyman resonance
line photons}
\label{ap:redist}

The redistribution function for an incoming Ly-$n^\prime$ and an
outgoing photon other than Ly-$n^\prime$ differs from the normal
redistribution function for scattering photons in which the outgoing
photon is the same as the ingoing photon (Ly-$n^\prime \rightarrow$
Ly-$n^\prime$). The principal difference arises from the re-emission
function following absorption. In the case of a preserved
Ly-$n^\prime$ photon, the outgoing frequency is the same as the
ingoing in the restframe of the atom. In the case of a degraded
photon, the outgoing frequency distribution is given by the
appropriate Lorentz profile \citep{1933Obs....56..291W}.

As an example, a scattering event is considered in which a
Ly-$n^\prime$ photon is degraded into a final Ly-$n$ photon, with two
intermediate transitions (the minimum required to produce a final
Lyman photon). The final photon need not be a Ly-$n$ photon; the
example only serves to illustrate the computations involved.

Accordingly, four levels are considered in the atom, labelled 0--3,
corresponding respectively to the ground state ($n=1$), the lowest
energy excited state, a higher intermediate state, and the $n^\prime$
energy level. The formalism and notation of
\citet{1933Obs....56..291W} are followed. The absorption profile for
an incoming photon of frequency $\xi^\prime$ in the frame of the atom
is
\begin{equation}
f(\xi^\prime)=\frac{1}{\pi}\frac{\delta_3}{(\xi^\prime-\nu_3)^2+\delta_3^2},
\label{eq:absphi}
\end{equation}
where the line-centre absorption frequency is $\nu_3$ and the total
decay width of the upper level is $\delta_3$. The decay from state 3
to state 2 involves recoil. Because the upper state was excited from a
definite energy level (the ground state), the upper energy level is
definite. As a consequence, only the uncertainty in the energy level
of state 2 will determine the width of the emission profile. Since the
atom will decay to the centre of energy level 2 on average, the
frequency of the emitted photons will peak at
$\xi=[h\xi^\prime-(\epsilon_2-\epsilon_0)]/h=\xi^\prime-\nu_2$, where
$\nu_2=(\epsilon_2-\epsilon_0)/h$. The probability density for
emitting a photon of frequency $\xi$ in the restframe of the atom is
then given by
\begin{equation}
p_{3,2}(\xi^\prime,\xi) = \frac{1}{\pi}
\frac{\gamma_2}{(\xi-\xi^\prime+\nu_2)^2+\gamma_2^2},
\label{eq:emiss32}
\end{equation}
where $\gamma_2$ is the decay width of state 2. The redistribution
function in the restframe of the atom is given by
$f(\xi^\prime)p_{3,2}(\xi^\prime,\xi)$. Note that if the incoming
photon is absorbed blueward (redward) of the line-centre, the emitted
photon frequency $\xi$ will also be shifted blueward (redward) of the
line centre frequency $\nu_{23}=(\epsilon_3-\epsilon_2)/h$.

For an atom moving at velocity ${\bf v}$ in the laboratory frame, the
redistribution function in terms of the incoming and outgoing photon
frequencies $\nu^\prime$ and $\nu$ in the laboratory frame becomes,
allowing for the Doppler shifting of the frequencies to first order in
$v/c$,
\begin{eqnarray}
&R_v(\nu^\prime,{\bf{\hat n}^\prime};\nu,{\bf\hat n})=
\frac{\delta_3\gamma_2}{\pi^2}\frac{1}{\left[\nu^\prime-\nu_3
\left(1+{\bf v}\cdot{\bf{\hat n}^\prime}/c\right)\right]^2
+\delta_3^2}\\
&\times\frac{1}{\left(\nu-\nu^\prime+\nu_3{\bf v}\cdot{\bf{\hat
      n}^\prime}/c - \nu_{23}{\bf v}\cdot{\bf\hat n}/c
+\Delta\nu_{\rm recoil}+\nu_2\right)^2 + \gamma_2^2},\nonumber
\label{eq:Rv23}
\end{eqnarray}
where the recoil term $\Delta\nu_{\rm
  recoil}=(h\nu_3\nu_{23}/m_ac^2)(1-{\bf\hat n}\cdot{\bf{\hat
    n}^\prime})$ has been included.

It is convenient to define the dimensionless frequencies
\begin{equation}
x=\frac{\nu-\nu_{23}}{\Delta\nu_D},\quad
x^\prime=\frac{\nu^\prime-\nu_3}{\Delta\nu_D},\quad{\rm
  with}\quad\Delta\nu_D=\nu_3\frac{b}{c},
\label{eq:xxpdef}
\end{equation}
where the frequencies are normalised by the same Doppler width. The
dimensionless decay widths $a_3=\gamma_3/\Delta\nu_D$ and
$a_2=\gamma_2/(\nu_{23}b/c)$ are also introduced, along with the
recoil parameter $\epsilon=h\nu_{23}/m_abc$ and the dimensionles
velocity ${\bf u}={\bf v}/b$. It is also convenient to introduce the
new coordinate system \citep{1962MNRAS.125...21H}
\begin{equation}
{\bf{\hat n}}_1=\gamma_{+}({\bf{\hat n}^\prime} + {\bf\hat n}),\quad
{\bf{\hat n}}_2=\gamma_{-}({\bf{\hat n}^\prime} - {\bf\hat n}),\quad
{\bf{\hat n}}_3={\bf\hat n}\times{\bf{\hat n}^\prime},
\label{eq:nbasis}
\end{equation}
where $\gamma_\pm = [2(1\pm\mu)]^{-1/2}$ and $\mu={\bf\hat
  n}\cdot{\bf{\hat n}^\prime}$. A dimensionless velocity-dependent
redistribution function is defined by $R_u(x^\prime,{\bf{\hat
    n}}^\prime;q,{\bf\hat n})=(\Delta\nu_D)^2R_v(\nu^\prime,{\bf{\hat
    n}^\prime};\nu,{\bf\hat n})$, where $q=x-x^\prime$. It is given by
\begin{eqnarray}
R_u(x^\prime,{\bf{\hat n}^\prime};q,{\bf\hat n})&=&
\left(\frac{a_3}{\pi}\right)\frac{1}{(x^\prime-\gamma_{+}^{-1}u_1/2-
\gamma_{-}^{-1}u_2/2)^2+a_3^2}\nonumber\\
&\times&\left(\frac{\nu_{23}a_2/\nu_3}{\pi}\right)
\Biggl\{\Biggl[q+\frac{1}{2}\gamma_{+}^{-1}u_1
\left(1-\frac{\nu_{23}}{\nu_3}\right)\nonumber\\
&+&\phantom{\Biggl[}\frac{1}{2}\gamma_{-}^{-1}u_2
\left(1+\frac{\nu_{23}}{\nu_3}\right)+\epsilon(1-\mu)\Biggr]^2\nonumber\\
&+&\phantom{\Biggl\{}\left(\frac{\nu_{23}}{\nu_3}a_2\right)^2\Biggr\}^{-1}.
\label{eq:Ru23}
\end{eqnarray}

It is convenient to perform the velocity average over a Maxwellian of
the redistribution function in Fourier space. The Fourier transform of
Eq.~(\ref{eq:Ru23}) is given by
\begin{eqnarray}
{\hat R}_u(\kappa,{\bf{\hat n}^\prime};\lambda,{\bf\hat
  n})&=&\int_{-\infty}^\infty
dx^\prime\,e^{ikx^\prime}\int_{-\infty}^\infty dq\,e^{i\lambda
q}R_u(x^\prime,{\bf{\hat n}^\prime};q,{\bf\hat n})\nonumber\\
&=&\exp\Biggl\{\frac{1}{2}i\gamma_{+}^{-1}u_1
\left(\kappa-\frac{\nu_2}{\nu_4}\lambda\right)\nonumber\\
&+&\frac{1}{2}i\gamma_{-}^{-1}u_2
\left[\kappa-\left(1+\frac{\nu_{23}}{\nu_3}\right)\lambda\right]\nonumber\\
\phantom{\Biggl\{}&-&i\lambda\epsilon(1-\mu)-a_3\vert\kappa\vert-
\frac{\nu_{23}}{\nu_3}a_2\vert\lambda\vert\Biggr\}.
\label{Ru23hat}
\end{eqnarray}
Averaging over a Maxwellian velocity distribution gives
\begin{eqnarray}
{\hat R}(\kappa,{\bf {\hat n}^\prime};\lambda,{\bf\hat n})&=&
\pi^{-3/2}\int d^3u\,e^{-u^2}
{\hat R}_u(\kappa,{\bf {\hat n}^\prime};\lambda,{\bf\hat n})\nonumber\\
&=&\exp\Biggl[-a_3\vert\kappa\vert-\frac{1}{4}\kappa^2-
\frac{\nu_{23}}{\nu_3}a_2\vert\lambda\vert\nonumber\\
&-&i\epsilon(1-\mu)\lambda+\frac{1}{2}\kappa\lambda-\frac{1}{4}
\left(1+\frac{\nu_{23}^2}{\nu_3^2}\right)\lambda^2\nonumber\\
&-&\phantom{\Biggl[}\frac{1}{2}
\frac{\nu_{23}}{\nu_3}(\kappa-\lambda)\lambda\mu\Biggr].
\label{eq:R23hat}
\end{eqnarray}

Moments of the dimensionless frequency shift are given by
\begin{equation}
a_n(x,\mu)=\int_{-\infty}^\infty dq\,q^nR(x^\prime,{\bf{\hat
    n}^\prime};q,{\bf\hat n}).
\label{eq:anxmu}
\end{equation}
The Fourier transforms of the moments may be expressed as
\begin{equation}
{\hat a}_n(\kappa,\mu)=\frac{\partial^n}{\partial(i\lambda)^n}
R(\kappa,{\bf{\hat n}^\prime};\lambda,{\bf\hat n})\Bigg\vert_{\lambda=0},
\label{eq:ankmu}
\end{equation}
noting that the factor $q^n$ corresponds to the $n^{\rm th}$
derivative with respect to $i\lambda$ in Fourier space. The
transformed moments are readily computed from
Eq.~(\ref{eq:R23hat}). Angle averaging the results and Fourier
transforming back to frequency space give for the first three moments,
\begin{eqnarray}
a_0(x)&=&\phi_V(a_3,x),\nonumber\\
a_1(x)&=&\frac{1}{2}\frac{d\phi_V(a_3,x)}{dx} -
\epsilon,\nonumber\\
a_2(x)&=&\frac{1}{2}\left(1+\frac{\nu_{23}^2}{\nu_3^2}\right)\phi_V(a_3,x)
-\epsilon\left(1+\frac{1}{3}\frac{\nu_{23}}{\nu_3}\right)
\frac{d\phi_V(a_3,x)}{dx}\nonumber\\
&+&\frac{1}{4}\left(1+\frac{1}{3}\frac{\nu_{23}^2}{\nu_3^2}\right)
\frac{d^2\phi_V(a_3,x)}{dx^2},
\label{eq:a012}
\end{eqnarray}
where only the first order terms in $\epsilon$ are retained. The first
order moment $a_1(x)$ is of the same form as for the scattering of
Ly-$n$ to Ly-$n$ \citep{Meiksin06}, and results in the heating rate
given by Eq.~(\ref{eq:Gn}). The Fourier transform of the
redistribution function averaged over angle is
\begin{eqnarray}
{\hat
  R}_{3,23}(\kappa,\lambda)&=&\frac{\sinh\left[\frac{1}{2}\frac{\nu_{23}}{\nu_3}
\lambda\left(\kappa-\lambda-2i\epsilon\frac{\nu_3}{\nu_{23}}\right)\right]}
{\frac{1}{2}\frac{\nu_{23}}{\nu_3}\lambda\left(\kappa-\lambda-2i\epsilon
\frac{\nu_3}{\nu_{23}}\right)}\nonumber\\
&\times&\exp\Biggl[-a_3\vert\kappa\vert-\frac{1}{4}\kappa^2-
\frac{\nu_{23}}{\nu_3}a_2\vert\lambda\vert\nonumber\\
&-&\phantom{\Biggl[}i\epsilon\lambda+\frac{1}{2}\kappa\lambda-
\frac{1}{4}\left(1+\frac{\nu_{23}^2}{\nu_3^2}\right)\lambda^2\Biggr],
\label{eq:Rkl323}
\end{eqnarray}
where the subscript on ${\hat R}_{3,23}(\kappa,\lambda)$ indicates the
scattering of a photon of frequency $\nu_3$ to $\nu_{23}$.

Similar redistribution functions may be defined for the decay product
photons of frequencies $\nu_{12}$ and $\nu_1$, however the analysis
become considerably more complicated as transitions between non-sharp
levels are involved. Following the discussion of
\citet{1933Obs....56..291W}, the frequency distributions may be
multiply peaked and correlated with the frequencies of previously
emitted photons. For example, for photons absorbed and emitted in the
Lorentz wings, the distribution of photons produced in the transition
from level 2 to level 1 could have two peaks. In one, the emitted
frequency would be correlated with the frequency $\nu_P$ of the photon
emitted in the transition $3\rightarrow2$, peaking at
$\nu^\prime-\nu_P-\nu_1$, followed by a transition $1\rightarrow0$
with the photon frequencies peaking at
$\nu_1=(\epsilon_1-\epsilon_0)/h$. A second peak would occur at the
frequency $\nu_{12}=(\epsilon_2-\epsilon_1)/h$, with the transition
$1\rightarrow0$ producing photons peaking in frequency at the
difference between $\nu_2$ and the frequency of the photon emitted in
the $2\rightarrow1$ transition. As the most probable frequency for the
final emitted Lyman photons (in this example) is independent of the
incoming frequency $\nu^\prime$, a good approximation to the
redistribution function Ly-$n^\prime$ to Ly-$n$ is
$R_{n^\prime,n}(\nu^\prime,\nu)=\varphi_V(a_{n^\prime},\nu^\prime)
\varphi_V(a_n,\nu)$.

\end{appendix}



\bibliographystyle{mn2e-eprint}
\bibliography{apj-jour,meiksin}

\label{lastpage}

\end{document}